\documentclass[useAMS,usenatbib]{mn2e}

\usepackage{url,times,graphicx,amsmath,amsfonts,amssymb,aas_macros,color,epsfig,epstopdf}

\newcommand{\ahf}{\textsc{AHF}}
\newcommand{\jumpd}{\textsc{JUMP-D}}

\newcommand{\subfind}{\textsc{SUBFIND}}

\newcommand{\rockstar}{\textsc{Rockstar}}

\newcommand{\stf}{\textsc{STF}}

\newcommand\Vmax{{\ifmmode{V_{\rm max}}\else{$V_{\rm max}$}\fi}}
\newcommand\Rmax{{\ifmmode{R_{\rm max}}\else{$R_{\rm max}$}\fi}}
\newcommand{\hMpc}{{\ifmmode{h^{-1}{\rm Mpc}}\else{$h^{-1}$Mpc}\fi}}
\newcommand{\hkpc}{{\ifmmode{h^{-1}{\rm kpc}}\else{$h^{-1}$kpc}\fi}}
\newcommand{\hMsun}{{\ifmmode{h^{-1}{\rm {M_{\odot}}}}\else{$h^{-1}{\rm{M_{\odot}}}$}\fi}}
\newcommand{\ltsima}{$\; \buildrel < \over \sim \;$}
\newcommand{\gtsima}{$\; \buildrel > \over \sim \;$}
\newcommand{\lsim}{\lower.5ex\hbox{\ltsima}}
\newcommand{\gsim}{\lower.5ex\hbox{\gtsima}}

\def\lesssim{\mathrel{\hbox{\rlap{\hbox{\lower4pt\hbox{$\sim$}}}\hbox{$<$}}}}
\def\gtrsim{\mathrel{\hbox{\rlap{\hbox{\lower4pt\hbox{$\sim$}}}\hbox{$>$}}}}

\newcommand{\Tab}[1]{Table~\ref{#1}}
\newcommand{\Sec}[1]{Section~\ref{#1}}

\newcommand{\Fig}[1]{Fig.~\ref{#1}}
\newcommand{\beq}{\begin{equation}}
\newcommand{\eeq}{\end{equation}}
\def\beqa{\begin{eqnarray}}
\def\eeqa{\end{eqnarray}}
\def\hMpc{$h^{-1}\,{\rm Mpc}$}
\def\hkpc{$h^{-1}\,{\rm kpc}$}

\title[Galaxies going MAD]
{Galaxies going MAD: The Galaxy-Finder Comparison Project}
\author[Knebe et. al]
       {Alexander Knebe$^{1}$\thanks{E-mail: alexander.knebe@uam.es}, 
        Noam I. Libeskind$^{2}$,
        Frazer Pearce$^{3}$,
        Peter Behroozi$^{4,5,6}$,\newauthor
        Javier Casado$^{1}$,
        Klaus Dolag$^{7,8}$,
        Rosa Dominguez-Tenreiro$^{1}$,
        Pascal Elahi$^{9}$,\newauthor
        Hanni Lux$^{3}$,
        Stuart I. Muldrew$^{3}$,
        Julian Onions$^{3}$\\
 $^{1}$Departamento de F\'isica Te\'orica, M\'odulo C-15, Facultad de Ciencias, Universidad Aut\'onoma de Madrid, 28049 Cantoblanco, Madrid, Spain\\
 $^{2}$Leibniz Institut f\"ur Astrophysik, An der Sternwarte 16, 14482 Potsdam, Germany\\
 $^{3}$School of Physics \& Astronomy, University of Nottingham, Nottingham, NG7 2RD, UK\\
 $^{4}$Kavli Institute for Particle Astrophysics and Cosmology, Stanford, CA 94309, USA\\
 $^{5}$Physics Department, Stanford University, Stanford, CA 94305, USA\\
 $^{6}$SLAC National Accelerator Laboratory, Menlo Park, CA 94025, USA\\
 $^{7}$Max-Planck Institut f\"{u}r Astrophysik, Karl-Schwarzschild Str. 1, D-85741 Garching, Germany\\
 $^{8}$University Observatory M\"unchen, Scheinerstr. 1, 81679, M\"unchen, Germany\\
 $^{9}$Key Laboratory for Research in Galaxies and Cosmology, Shanghai Astronomical Observatory, Shanghai 200030, China\\
}

\begin{document}

\date{Accepted XXXX . Received XXXX; in original form XXXX}

\pagerange{\pageref{firstpage}--\pageref{lastpage}} \pubyear{2010}

\maketitle

\label{firstpage}


\begin{abstract}
With the ever increasing size and complexity of fully self-consistent
simulations of galaxy formation within the framework of the cosmic
web, the demands upon object finders for these simulations has
simultaneously grown. To this extent we initiated the Halo Finder
Comparison Project that gathered together all the experts in the field
and has so far led to two comparison papers, one for dark
matter field haloes \citep{Knebe11}, and one for dark matter subhaloes
\citep{Onions12}. However, as state-of-the-art simulation codes are
perfectly capable of not only following the formation and evolution of
dark matter but also account for baryonic physics, i.e. gas
hydrodynamics, star formation, stellar feedback, etc., object finders
should also be capable of taking these additional physical processes
into consideration.  Here we report -- for the first time -- on a
comparison of codes as applied to the Constrained Local UniversE
Simulation (CLUES) of the formation of the Local Group which
incorporates much of the physics relevant for galaxy formation. We
compare both the properties of the three main galaxies in the
simulation (representing the Milky Way, Andromeda, and M33) as
well as their satellite populations for a variety of halo finders
ranging from phase-space to velocity-space to spherical overdensity
based codes, including also a mere baryonic object finder. We obtain agreement amongst codes comparable to (if not
better than) our previous comparisons -- at least for the total, dark,
and stellar components of the objects. However, the diffuse gas
content of the haloes shows great disparity, especially for low-mass
satellite galaxies. This is primarily due to differences in the
treatment of the thermal energy during the unbinding procedure. We
acknowledge that the handling of gas in halo finders is
something that needs to be dealt with carefully, and the precise
treatment may depend sensitively upon the scientific problem being
studied.

\end{abstract}
\noindent
\begin{keywords}
  methods: $N$-body simulations -- galaxies: haloes -- galaxies:
  evolution -- cosmology: theory -- dark matter
\end{keywords}

\section{Introduction} \label{sec:introduction}
In a series of precursory papers \citep[][see also Knebe et al., in
prep.]{Knebe11,Onions12} we have compared and quantified the
differences arising from the application of various halo finders to
the same (dark matter only) data sets. The motivation for such
comparisons is obvious: while past decades have seen great
strides to better understand disparities in cosmological simulations
stemming from different gravity (and hydrodynamics) solvers
\citep[e.g.][]{Frenk99, Knebe00, OShea05, Agertz07, Heitmann08cccp,
Tasker08,Robertson10,Springel10,Vazza11} it was about time to initiate
a similar project for object finders. To put it in a provocative way,
if you are not only interested in the matter distribution in the universe,
any simulation is only as good as the halo catalogue derived from it,
especially at a time when we will have access to extremely large
redshift surveys (just to name a few, BOSS, WiggleZ, eBOSS, BigBOSS,
DESpec, PanSTARRS, DES, HSC, Euclid, WFIRST, etc.) against which the
next generation of cosmological simulations (and the haloes/galaxies
found in them) should and will be compared against.

However, all our previous comparisons solely focused on dark matter
haloes and left aside any possible influence of the baryons. In this
work we extend the comparison project to a cosmological simulation
that not only models gravity, but simultaneously follows the evolution
of the baryonic material by incorporating a self-consistent solution
to the hydrodynamics. The simulation further includes star formation
and stellar feedback, all explained in more detail in the subsequent
\Sec{sec:data}. Please note that we are not trying to address the
question of (disk) galaxy formation in this work, but rather aim at
using one of these simulations to verify whether a halo finder applied
to it will find the same (galactic) objects as any other halo
finder. In particular, we will study both the larger (disk) galaxies
as well as their respective satellite populations. To this extent we
use a constrained zoom simulation of the Local Group of
galaxies consisting of objects akin to the Milky Way (MW), the
Andromeda galaxy (M31), and M33. This data forms part of the
Constrained Local Universe (CLUES)
project.\footnote{\texttt{http://www.clues-project.org}}

The questions to be addressed are plain and simple: while it is well
established that baryonic physics alters the particulars of dark matter
haloes and subhalo populations
\citep{Blumenthal86,Tissera98,Maccio06,Onorbe07,Romano-Diaz08,Romano-Diaz09,Libeskind10,Duffy10,Sommer-Larsen10,Romano-Diaz10,Schewtschenko11,DiCintio11,Governato12,Zolotov12,Brooks12,DiCintio12},
there remains the question of how this will influence the performance
of an object finder. Note that cold baryonic material tends to cluster
more strongly than its dark matter counterpart due to the dissipative
nature of the cooling process allowing for the disposal of energy via
radiation. While star particles -- generated on-the-fly during the
simulation according to a physically motivated prescription that
transforms cold gas into stars -- only feel gravitational forces
(like the dark matter particles), they nevertheless originate from
compact, condensed regions. Additionally the gas particles themselves
carry not only kinetic but also thermal energy giving rise to thermal
pressure in the medium (specified by the adopted equation of
state). At present, halo finders deal with these subtleties
differently, with some properly accounting for the gas pressure and
others ignoring it. Therefore, the fundamental question is,
what will different halo/galaxy finders find when it comes to
simulations including baryonic physics?

Here we compare a range of different (halo) finding techniques ranging from phase-space to
velocity-space to spherical overdensity based codes, but also including a mere baryonic object finder.
This naturally raises the question about the right approach to actually find galaxies. Should one
hunt for dark matter haloes first and then characterize the galaxy residing in it? Or is it more
reasonable to initially locate galaxies and then identify their surrounding dark matter haloes? Our suite
of object finders contains both methods of operation and we will hence learn about these different strategies, too.

\section{The Data} \label{sec:data}
We use the same simulation already presented in
\citet{Libeskind10,Gottloeber10,Libeskind10infall,Knebe10a,Libeskind11,Knebe11a,DiCintio11,Dayal12,DiCintio12}
and refer the reader to these papers for a more exhaustive discussion
and presentation of our constrained simulations of the Local Group
that form part of the aforementioned CLUES project. We
briefly summarise their main properties here for clarity.

\subsection{Constrained Simulations of the Local Group}
\label{sec:localgroup}
We choose to run our simulations using standard $\Lambda$CDM initial
conditions, that assume a WMAP3 cosmology \citep{Spergel07},
i.e. $\Omega_m = 0.24$, $\Omega_{b} = 0.042$, $\Omega_{\Lambda} =
0.76$. We use a normalisation of $\sigma_8 = 0.73$ and a $n=0.95$
slope of the power spectrum. We used the treePM-SPH code
\texttt{GADGET2} \citep{Springel05} to simulate the evolution of a
cosmological box with side length of $L_{\rm box}=64 h^{-1} \rm
Mpc$. Within this box we identified (in a lower-resolution run
utilizing $1024^3$ particles) the position of a model local group that
closely resembles the real Local Group
\citep[cf.][]{Libeskind10}. This Local Group has then been re-sampled
with 64 times higher mass resolution in a region of $2 h^{-1} \rm Mpc$
about its centre giving a nominal resolution equivalent to $4096^3$
particles, resulting in a mass resolution of $m_{\rm DM}=2.5\times
10^{5}$\hMsun\ for the dark matter and $m_{\rm gas}=4.4\times
10^4$\hMsun\ for the gas particles. The force resolution is
$0.15$\hkpc. 

For this particular study we focus on the gas dynamical smoothed
particle hydrodynamics (SPH) simulation, in which we follow the
feedback and star formation rules of \cite{Springel03}: the
interstellar medium (ISM) is modelled as a two phase medium composed of
hot ambient gas and cold gas clouds in pressure equilibrium. The
thermodynamic properties of the gas are computed in the presence of a
uniform but evolving ultra-violet cosmic background generated from
QSOs and AGNs and switched on at $z=6$ \citep{Haardt96}.  Cooling
rates are calculated from a mixture of a primordial plasma
composition. No metal dependent cooling is assumed, although the gas
is metal enriched due to supernovae explosions. Molecular cooling
below $10^{4} {\rm K}$ is also ignored. Cold gas cloud formation by
thermal instability, star formation, the evaporation of gas clouds,
and the heating of ambient gas by supernova driven winds are assumed
to all occur simultaneously.

\begin{figure}
  \includegraphics[width=\columnwidth]{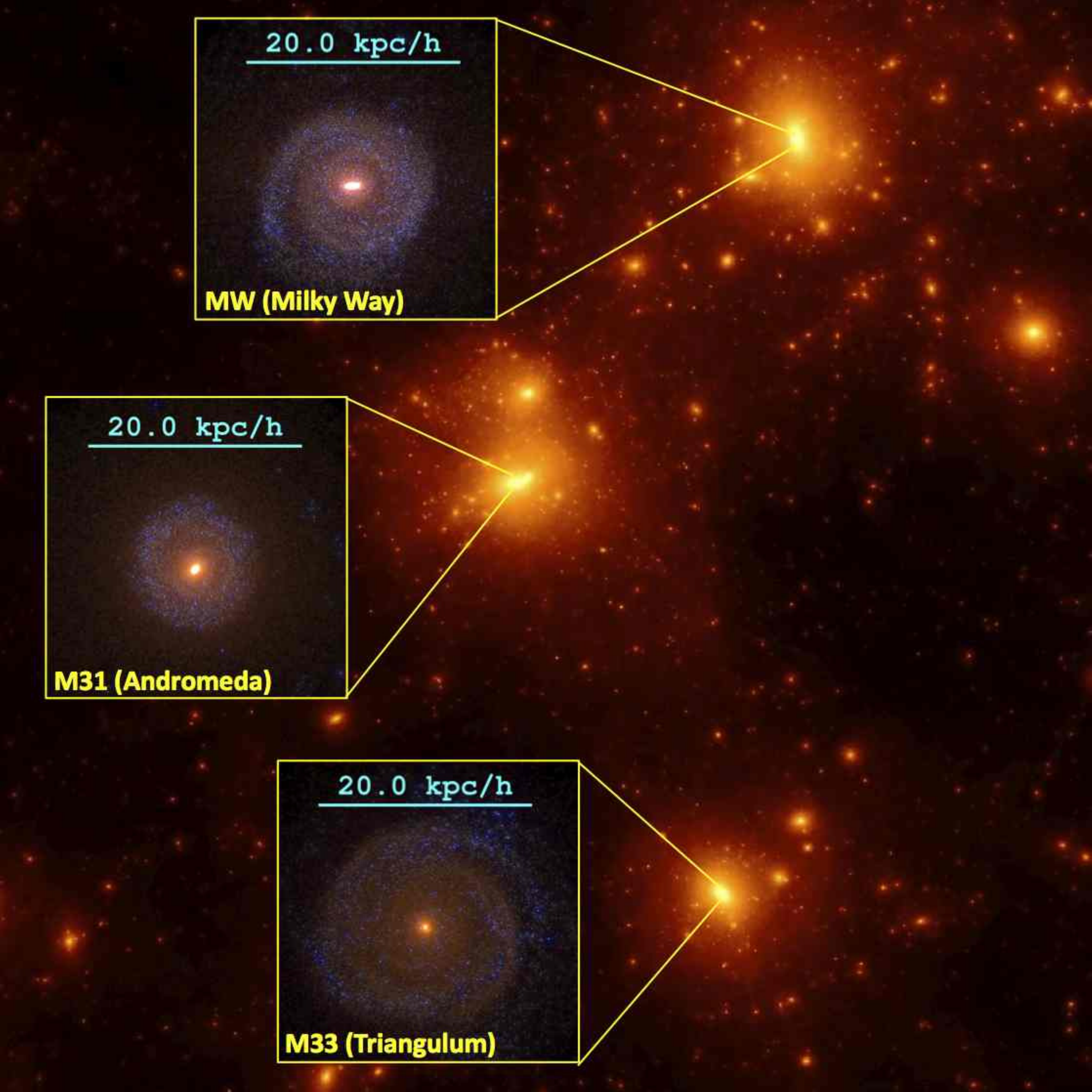}
  \caption{A graphical representation of the CLUES simulation data showing the three main galaxies considered in this study, i.e. M31, the MW and M33. In \Sec{sec:galaxycomparison} the properties of
  those three main galaxies will be compared whereas
  Sections~\ref{sec:satellitecomparison} and~\ref{sec:crosscomparison}
  focus on the satellite populations of M31 and the MW. \textit{Image courtesy CLUES collaboration.}}
\label{fig:localgroup}
\end{figure}

To give a visual impression of our simulated Local Group we present in \Fig{fig:localgroup} a colour-coded density map of it as well as a zoom of the stellar disks of three main galaxies considered in this study. For more details we refer the reader to the CLUES web
site or the aforementioned papers. But please note that the actual
simulation is not up for debate here: the prime objective of this work
is to compare objects finders and hence any reasonable simulation
of galaxy formation might serve this purpose.

\subsection{Lighting up (Sub-)Haloes}
\label{sec:lightingup}

The stellar population synthesis model STARDUST \citep[see][and
references therein for a detailed description]{Devriendt99} has been
used to derive luminosities from the stars formed in our
simulation. This model computes the spectral energy distribution from
the far-UV to the radio, for an instantaneous starburst of a given
mass, age and metalicity. The stellar contribution to the total flux
is calculated assuming a Kennicutt initial mass function
\citep{Kennicutt98}.

\section{The Codes} \label{sec:codes}
Before (briefly) presenting the particulars of the codes, it is
important to state that all of the participating codes are at various
stages with respect to their capabilities of handling simulation data
that includes not only (equal) mass dark matter particles, but also
star and gas particles. For instance, \subfind\, with its careful
treatment of the gas thermal properties, is the most advanced
with respect to the simulation data at hand, whereas on the other hand
\rockstar\ required substantial post-processing as it treats all
particles on a equal footing at present.


\subsection{\ahf\ (Knebe)}
The halo finder \ahf\footnote{\ahf\ is freely available from
\texttt{http://www.popia.ft.uam.es/AHF}} (\textsc{Amiga} Halo Finder)
is a spherical overdensity finder that identifies (isolated and
sub-)haloes as described in \citet{Gill04a} as well as
\citet{Knollmann09}. The initial particle lists are obtained by a
rather elaborate scheme: for each subhalo the distance to its nearest
more massive (sub-)halo is calculated and all particles within a
sphere of radius half this distance are considered prospective subhalo
constituents. This list is then pruned by an iterative unbinding
procedure using the (fixed) subhalo centre as given by the local
density peak determined from an adaptive mesh refinement
hierarchy. For more details we refer the reader to the aforementioned code
description papers as well as the online documentation.

\paragraph*{Gas Treatment} For the data given here \ahf\ version v1.0-021 was run in two configurations: one where the
thermal energy of the gas entered the unbinding procedure via

\begin{equation}
 e_{\rm gas} = \phi + \frac{1}{2} v^2 + u
\end{equation}

\noindent
where $e_{\rm gas}$ is the total specific energy (required to be
negative for bound particles), $\phi$ the gravitational potential, $v$
the gas particle's velocity and $u=\frac{3}{2} \frac{k_B}{m} T$ its
thermal energy, and a second one where this feature was switched off
(i.e. $u=0$).

\subsection{\jumpd\ (Casado \& Dominguez-Tenreiro)} \label{sec:jumpd}
\jumpd\ is a galaxy finder and \textit{not} a sub-\textit{halo} finder and hence is treated differently than the other finders in this work. It aims at finding and measuring central and satellite galaxies within given host haloes, i.e. \textit{baryonic} substructure objects within a sphere of given radius $R_{\rm lim}$ about the centre of the host. To this extent the stellar and gas mass profiles are searched for jumps (and hence the name) in the three-dimensional cumulative mass profiles  from the host halo centre out to the limiting radius $R_{\rm lim}$ (i.e. usually the host halo's virial radius). The jump detection criterion is based on the detection of changes in the first and second derivatives of the respective mass profiles in the $r, \theta$ and $\phi$ variables at the substructure locations corresponding to the humps they cause. For the stellar object, the jump in the stellar mass profile is used as a first satellite detection (i.e., location and velocity),  that is later on refined by searching for maxima in 6-dimensional phase-space within 
 an allowance region about that first center, returning  the object stellar sizes $r_{\rm star}$ as well. The jumps  in the gas profile are then matched to the stellar objects and gas particles inside a spherical region defined by the radial extend of the gas jump ($r_{\rm gas}$) are then  associated to  the stellar object. Note that for the detection of the jumps only cold gas is considered.

Please note that the approach of \jumpd\ is substantially different to halo finders in general. The code \textit{only} locates a baryonic object ("galaxy") without considering the dark matter. To this extent \jumpd\ cannot be subjected to the common post-processing pipeline (cf. \Sec{sec:commonpipeline}) when it comes to subhaloes as that pipeline heavily relies on the embedding of satellite galaxies within dark matter subhaloes. Therefore, \jumpd\ only contributes to those plots where the identities of such dark matter subhalo particles is not required. The situation is different for field/host haloes where it is more straight forward to identify the surrounding dark matter halo and its virial radius. In that regards, \jumpd\ has been passed through the pipeline for the results presented in \Sec{sec:galaxycomparison}.

\paragraph*{Gas Treatment} For the satellite galaxies we take advantage of the bi-phase nature of the gas in this particular simulation, and just consider the cold gas when detecting jumps in the gas profile. To determine the cut in internal energy separating both the cold and hot gas regimes, the thermal energy distribution of gas particles within the host virial radii is studied and its minimum  is used to separate the two phases and reject the hot gas from the object, respectively. However, this is only applied to the satellite galaxies and not the three main objects where all gas irrespective of its phase is added.

\subsection{\rockstar\ (Behroozi)}
\rockstar\ (Robust Overdensity Calculation using K-Space Topologically
Adaptive Refinement) is a phase-space halo finder designed to maximise
halo consistency across timesteps \citep{Behroozi12}.  The algorithm
first selects particle groups with a 3D Friends-of-Friends (FOF)
variant with a very large linking length ($b = 0.28$). For each main
FOF group, Rockstar builds a hierarchy of FOF subgroups in phase-space
by progressively and adaptively reducing the linking length, so that a
tunable fraction (70 per cent, for this analysis) of the particles are
captured within each subgroup as compared to the immediate parent
group.  When this is complete, Rockstar converts FOF subgroups into
seed haloes beginning at the deepest level of the hierarchy. If a
particular group has multiple subgroups, then particles are assigned
to the subgroup seed haloes based on their phase-space proximity. This
process is repeated at all levels of the hierarchy until all particles
in the base FOF group have been assigned to haloes. Unbinding is
performed using the full particle potentials; halo centres and
velocities are calculated in a small region close to the phase-space
density maximum.  Rockstar is a massively parallel code (hybrid
OpenMP/MPI style).

\paragraph*{Gas Treatment} As already mentioned, initially \rockstar\ 
provided a particle ID list that was based upon an analysis that
treated each particle equally (without even taking into account the
difference in mass between the phases). In the common post-processing
stage (see \Sec{sec:commonpipeline} below for more details) an
unbinding procedure was then applied outside of \rockstar. However,
this unbinding (deliberately) did not take into account the thermal
energy of the gas, but only considered the kinetic and potential
energy of all the particles.

\begin{table*}
  \caption{Some general properties of the three main galaxies. Masses
  are measured in $10^{11}$\hMsun, velocities in km/sec, magnitudes
  are the Johnson V-band magnitude, and the baryon fractions are given
  as $f_X=M_X/M_{\rm tot}$ where $f_X$ can be either $f_g$ (gas mass
  fraction) or $f_s$ (stellar mass fraction). Note that the total
  baryon fraction is the sum $f_b=f_g+f_s$.}
\label{tab:generalproperties}
\begin{center}
\begin{tabular}{lccccccccccccccccccc}
\hline
 code		& \multicolumn{3}{c}{Mass}	&& \multicolumn{3}{c}{\Vmax}	&& \multicolumn{3}{c}{Johnson $M_V$}	&& \multicolumn{3}{c}{gas mass fraction $f_g$}	&& \multicolumn{3}{c}{stellar mass fraction $f_s$}\\
 	 		& MW 	& M31	& M33 	&& MW	& M31 	& M33	&& MW	& M31	& M33	&& MW	& M31	& M33	&& MW	& M31	& M33\\
\hline
\ahf			& 3.4		& 4.3		& 1.7		&& 124	& 125	& 119	&& -20.3	& -21.0	& -19.5	&& 0.043	& 0.078	& 0.066	&& 0.035	& 0.033	& 0.037\\
\jumpd		& 3.4		& 4.2		& 1.7		&& 124	& 125	& 118	&& -20.3	& -20.9	& -19.5	&& 0.043	& 0.077	& 0.066	&& 0.035	& 0.034	& 0.037\\
\rockstar		& 3.4		& 4.3		& 1.7		&& 124	& 125	& 119	&& -20.3	& -21.0	& -19.5	&& 0.043	& 0.078	& 0.066	&& 0.035	& 0.033	& 0.037\\
\stf			& 3.1		& 3.3		& 1.6		&& 124	& 121	& 119	&& -20.3	& -20.6	& -19.5	&& 0.043	& 0.079	& 0.067	&& 0.036	& 0.032	& 0.037\\
\subfind		& 3.2		& 3.5		&1.6		&& 124	& 123	& 119	&& -20.2	& -20.6	& -19.5	&& 0.042	& 0.069	& 0.058	&& 0.035	& 0.031	& 0.037\\
\hline
\end{tabular}
\end{center}
\end{table*}

\subsection{\stf\ (Elahi)}
The STructure Finder is a parallel hybrid OpenMP/MPI code
\citep[\stf]{Elahi11} that identifies objects by utilizing the fact
that dynamically distinct substructures in a halo will have a {\em
local} velocity distribution that differs significantly from the mean,
{\em i.e.} smooth background halo. This method consists of two main
steps, identifying particles that appear dynamically distinct and
linking this outlier population using a Friends-of-Friends-like
approach. Since this approach is capable of not only finding
subhaloes, but tidal streams surrounding subhaloes as well as tidal
streams from completely disrupted subhaloes, for this analysis we also
ensure that a group is self-bound. Particles which are gravitationally
unbound to a candidate subhalo are discarded until a fully self-bound
object is obtained or the object consists of fewer than 20 particles,
at which point the group is removed.

\paragraph*{Gas Treatment} For this study, we treat each particle type 
separately, first identifying dark matter (sub)haloes. Gas and star
particles are then associated with the closest dark matter particle in
phase-space belonging to a (sub)halo. Further, in order to highlight
differences in how particles can be treated, we \textit{do not} pass
baryonic particles through an unbinding routine. Part of the
motivation for this comes from observational studies. It is quite
difficult to determine from observations alone whether gas or stellar
structures are bound to a galaxy. Instead, spatially coincident gas
and stellar structures lying within the same line-of-sight velocity
window are generally assumed to be dynamically related. Hence in this
spirit, we do not attempt to determine whether a gas or star particle
is bound to a dark matter (sub)halo, nor do we attempt to account for
the thermal properties of the gas.

We note that multiple masses do not effect the local velocity
distribution function estimate though they do affect the background
velocity distribution estimate, specifically the centre-of-mass of the
coarse grain cells used to estimate the local velocity
distribution. 

\subsection{\subfind\ (Dolag \& Springel)}
\subfind\ identifies substructures as locally overdense, gravitationally
bound groups of particles. Starting with a halo identified through the
Friends-of-Friends algorithm, a local density is estimated for each particle
with adaptive kernel estimation using a prescribed number of smoothing
neighbours. Starting from isolated density peaks, additional particles are added
in sequence of decreasing density. Whenever a saddle point in the global density
field is reached that connects two disjoint overdense regions, the smaller
structure is treated as a substructure candidate, followed by merging the two
regions. All substructure candidates are subjected to an iterative unbinding
procedure with a tree-based calculation of the potential. The \subfind\ algorithm
is discussed in full in \citet{Springel01subfind} and its extension to dissipative
hydrodynamical simulations that include star formation in \citet{Dolag09}.

\paragraph*{Gas Treatment} Within \subfind\ the total density at the
position of each particle is estimated by summing up the contribution
of all the different particle species. In contrast to applying the SPH
formalism to all particles at the same time, this allows a much more
fine structured density field to be obtained if the particles of the
different species have significantly different spatial distributions,
as is usually the case for star particles.  We use all the particles
initially belonging to the substructure candidate in order to evaluate
the gravitational potential. For gas particles, we also take the
internal thermal energy into account in the gravitational unbinding
procedure. If the number of bound particles left is larger than 20 and
the object contains at least one star or one dark matter particle, we
register the substructure as genuine sub-halo.

\begin{figure*}
  \includegraphics[width=2.15\columnwidth]{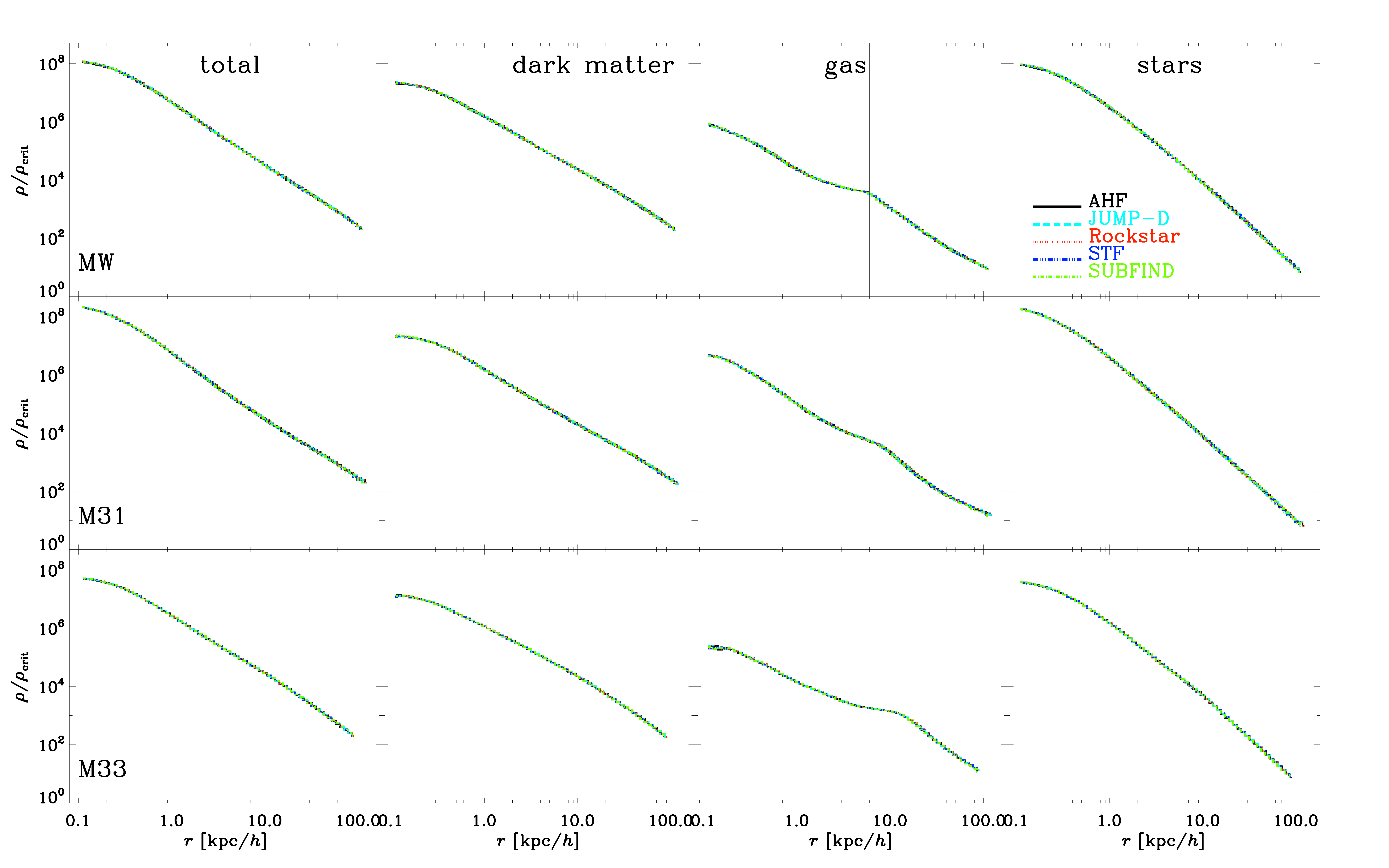}
  \caption{Density profile of the total (first column), dark (second
  column), gas (third column), and stellar (fourth column) matter for
  M31 (upper row), the MW (middle row), and M33 (bottom row). The
  vertical line (coinciding with the 'kink') in the gas profile at 8,
  6, and 10\hkpc, respectively, indicates the transition from the
  central baryon dominated region to the halo and will be used for the
  bulge-disk decomposition in \Sec{sec:bdecomp}.}
\label{fig:DensProfile}
\end{figure*}

\begin{figure*}
  \includegraphics[width=2.15\columnwidth]{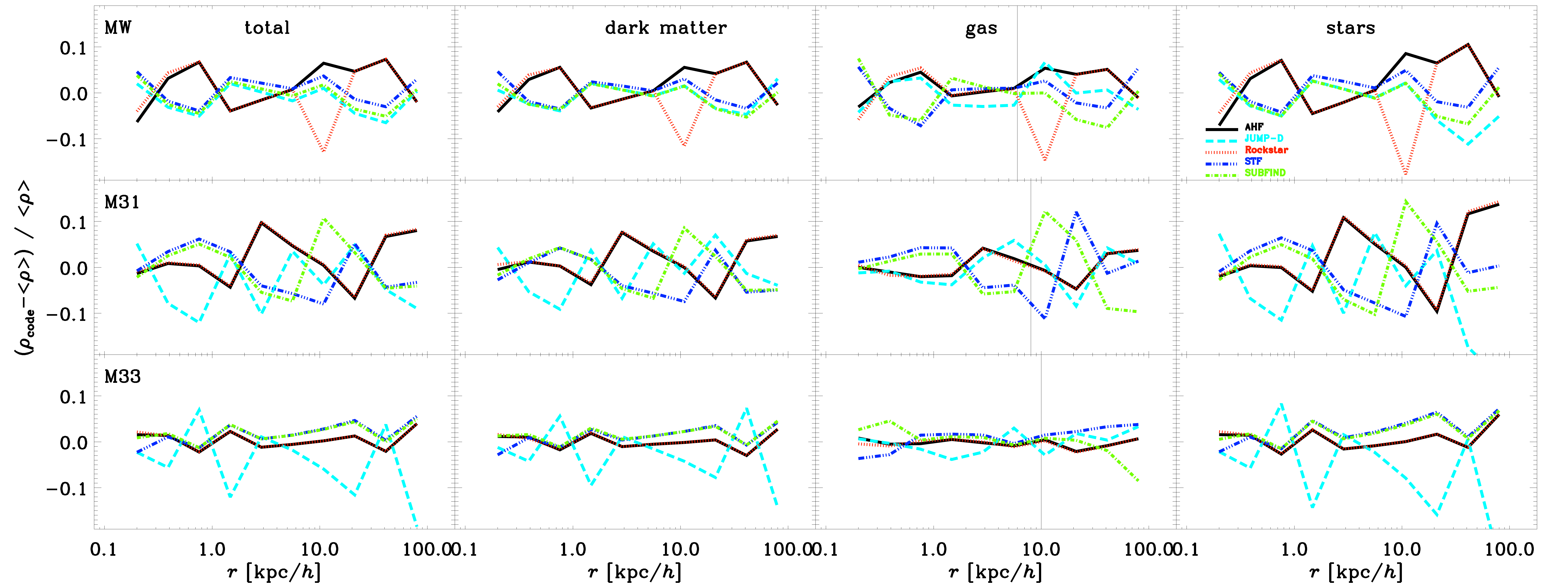}
  \caption{Same as \Fig{fig:DensProfile} but this time showing the relative difference to the mean of all density profiles.}
\label{fig:DensProfileRatio}
\end{figure*}

\subsection{Common Post-Processing Pipeline} \label{sec:commonpipeline}
As was the case for the previous comparison projects we again
subjected all halo finder (excpet for \jumpd\ satellite galaxies) results to a common post-processing
pipeline. Code representatives were simply asked to return the
particle IDs of those particles they considered to belong to a given
object (may that be a host or sub-halo). To this extent, halo
positions are iteratively determined centre-of-masses using the
innermost 5 per cent of particles (limiting the positional shift per
iteration to the force resolution of the simulation, i.e. 0.15\hkpc),
the bulk velocity is the mean velocity of all particles, the mass
corresponds to $M_{200}$ (i.e. the mass enclosed in a sphere so that
the mean density inside the sphere equals 200 times the critical
density of the Universe), and \Vmax\ is the peak value of the circular
velocity curve. This approach entails that any scatter reported here
will be a lower limit as all halo properties are defined in an
identical manner. The star particles and their metalicities and ages
are used to calculate the luminosities and magnitudes
(cf. \Sec{sec:lightingup} for the model). Note that we primarily use
magnitude throughout this work, liberally referring to it as
"luminosity" in places.

We also need to mention that the mode of operation of this
pipeline biases results towards galaxies residing in dark matter
haloes: both the definition of the object's edge and the fact that
we require \Vmax\ to be calculated relies on the fact that the 
bound dark matter particles within such a subhalo
have been identified. Therefore -- as mentioned before -- the \jumpd\  
satellite outputs are not best suited to be passed through this pipeline.

\section{Galaxy  Comparison} \label{sec:galaxycomparison}
The present work would not rightfully be called a "Galaxy Comparison
Project" if we were not to study the properties of the (baryonic
component of the) dominant galaxies in our constrained Local Group
simulation, i.e. the MW, M31, and M33. We therefore begin by comparing the particulars of these
objects.

\subsection{General Properties}\label{sec:generalproperties}
In \Tab{tab:generalproperties} we provide a summary of
the most fundamental properties of the three main galaxies in our
simulation, i.e. we list the mass, \Vmax, luminosity (as characterised
by the Johnson V-band magnitude), and baryon fractions $f_X=M_X/M_{\rm
tot}$ where $f_X$ can be either $f_g$ (gas mass fraction) or $f_s$
(stellar mass fraction). Note that the total baryon fraction is the
sum $f_b=f_g+f_s$. There is a rather excellent agreement across
the halo finders, with the scatter in basically any quantity
being smaller than 10 per cent, which was the deviation found
in our previous dark matter only comparisons. The remaining
differences seen in the mass are readily explained by the fact that
some codes (e.g. \ahf) consider subhaloes to contribute to the mass of
their respective host while others (e.g. \stf\ and \subfind) exclude
the subhalo masses.

\subsection{Mass Distributions}\label{sec:DensProfile}
To explore any differences in more detail we now decompose the mass
profiles of the three galaxies. The result is presented in
\Fig{fig:DensProfile} where we show the total matter density profile
(left column) and its breakdown into the various components,
i.e. dark matter (2nd column), gas (3rd column), and stars (rightmost
column). The rows represent the MW (upper row), M31 (middle row), and
M33 (bottom row).

All finders agree outstandingly well
for all three components (and hence the total matter, too). Despite
their different treatment of the gas thermal energy, the resulting
mass distributions throughout the three galaxies remain essentially
indistinguishable. When comparing the profiles to the numbers in
\Tab{tab:generalproperties} one may wonder about M31 and why it is
that while the masses differ by 30 per cent the density profiles agree
so remarkably well. This is readily explained by differences in the
virial radii of approximately 10 per cent, something not
clearly visible in \Fig{fig:DensProfile} due to the logarithmic
$x$-axis. In that regard, note that all the curves in the figure
terminate at the respective virial radius and go slighter closer to the centre
than the nominal force resolution of the simulation,
i.e. $0.15$\hkpc. These variations in mass and radius are partly due to 
different treatment of subhalo masses: some codes include substructure
in the host mass (e.g. \ahf) whereas other do not (e.g. \subfind). And this
is particularly important for the simulated M31 as it contains one
large group of satellites contributing approx. 15 per cent of the total host mass
\citep[cf. Sec.~4.2.1 in][where this particular object is discussed in detail]{Klimentowski10}.
However, the mass and radii differences are also affected by disparate
particle collection and unbinding procedures (Knebe et al., in preparation) accounting
for the remaining variations across codes.

Another interesting aspect of \Fig{fig:DensProfile} is the kink in the
gas density profile apparent at approximately 6-10\hkpc, corresponding
to roughly 10 per cent of the virial radius (with no difference across
finders again). This peak represents the transition from the central
baryon dominated region to the halo and will be used in the following
Sub-Section for the bulge-disk decomposition.

\begin{figure*}
  \includegraphics[width=40pc]{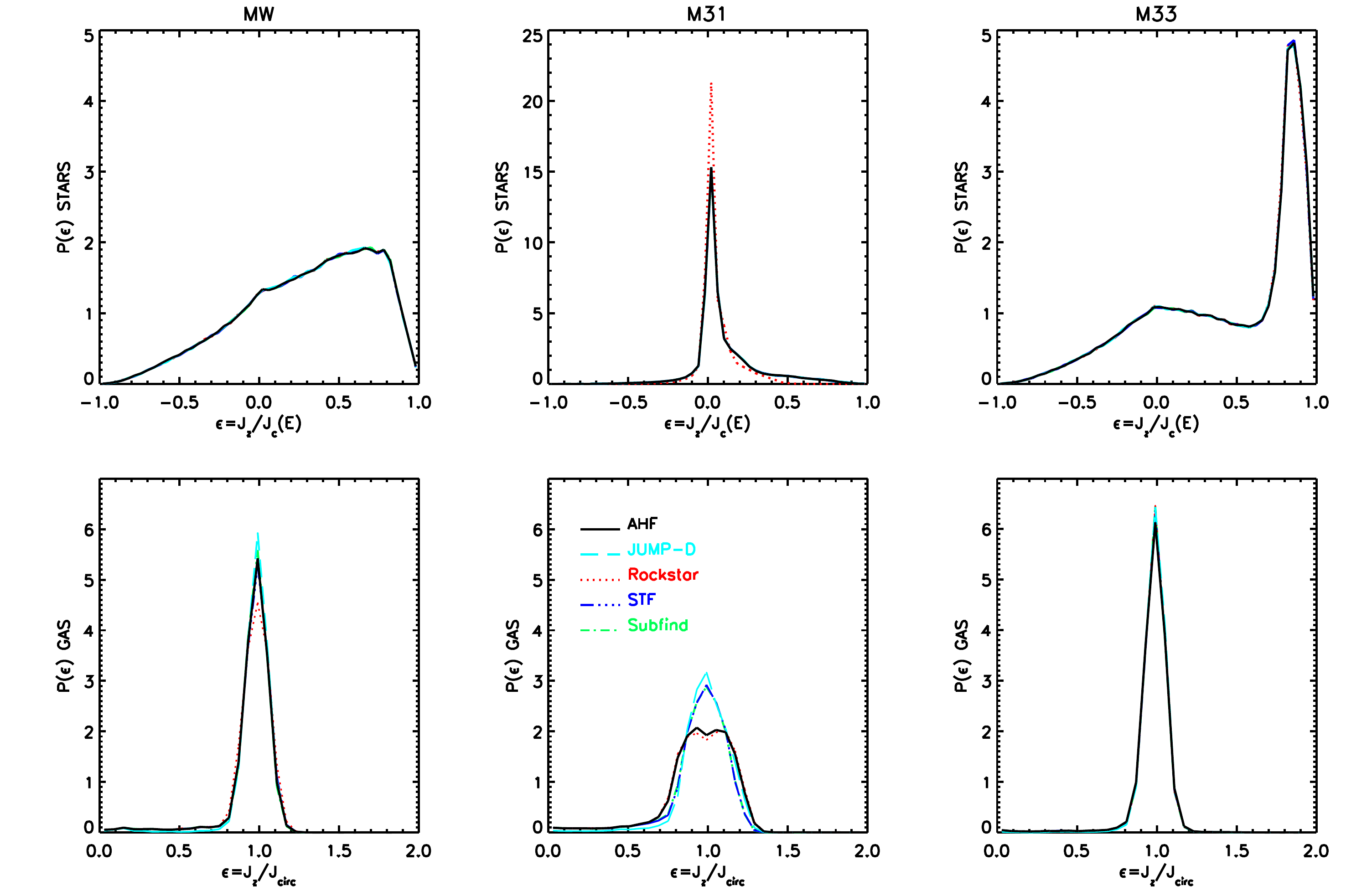}
  \caption{A dynamical bulge - disc decomposition for stars (top) and
  gas (bottom) particles in the MW (left), M31 (middle) and M33
  (right). We compare each star particle's angular momentum in the
  $z$-direction to the angular momentum of a circular orbit of the
  same energy and show on the $y$-axis the probability density
  distribution $P(\epsilon)$, for a given $\epsilon=J_{z} / J_{\rm
  c}(E)$. For gas particles we compare the angular momentum in the
  $z$-direction with the angular momentum of a circular orbit at that
  radius and show on the $y$-axis the probability density distribution
  $P(\epsilon)$, for a given $\epsilon=J_{z} / J_{\rm circ}$.}
\label{fig:bdecomp-comp}
\end{figure*}

To better gauge the differences across finders we additionally plot
in \Fig{fig:DensProfileRatio} the fractional differences between the
profile of each finder and the respective mean profile when averaging over
all codes. This figure shows that the residual scatter of the density profiles
is at most just 10 per cent about the mean.

\subsection{Dynamical Bulge--Disk Decomposition}\label{sec:bdecomp}
A dynamical bulge--disk decomposition for all three central galaxies
in our simulation has been performed. We decompose the galaxies using
two different techniques, one for the gas and one for the stars.

For the star particles we apply the method proposed by
\citet{Abadi03b}, used by \citet{Okamoto05}, and improved upon in
\citet{Domenech-Moral12}. The method is to essentially compare the
$z$-component of the angular momentum ($J_{z}$) of each particle with
the angular momentum of a circular orbit with the same energy, $J_{\rm
c}(E)$. The first step is to thus define a $z$-direction which, as
suggested by \cite{Abadi03b} may be chosen to be the total (specific)
angular momentum of all the star particles inside the "luminous
radius" of the galaxy (defined here as within 6, 8, and 10\hkpc\ for
the MW, M31, and M33, respectively; see the kink in the gas profiles
visible in \Fig{fig:DensProfile}). The choice of a $z$-direction in
this manner ensures that the majority of the particles will have
positive values of $J_{z}$. The potential and kinetic energy for each
star and gas particle within this luminous radius of the galaxy is
then calculated. Since circular orbits maximise angular momentum, we
assume $J_{\rm c}(E)$ for a particle of energy $E$ is simply the
maximum value of $J_{z}$ of all particles with that energy. Note that
the limits of $\epsilon=J_{z} / J_{\rm c}(E)$, i.e. -1 and 1,
correspond to counter- and co-rotating particles on circular
orbits. Spheroidal systems with very little net rotation are thus
represented by a Gaussian centred on 0.  A fully rotationally
supported thin disk on the other hand is represented by values of
$J_{z} / J_{\rm c}(E)\approx 1$. Thick disks not fully supported by
their rotation can be seen as distributions centred on positive
values such as $J_{z} / J_{\rm c}(E)\approx 0.5$ \citep{Abadi03b}.

The resulting distributions can be viewed in the upper panels of
\Fig{fig:bdecomp-comp}. All three galaxies contain both co- and
counter rotating populations. The MW comprises a bulge and a
(co-rotating) thick disk that peaks at $\epsilon\approx0.8$. No thin
disc is seen here, although owing to their relatively high values of
$\epsilon$, many particles in the thick disc are kinematically
cold. M31 on the other hand appears as a featureless spheroid whose
net angular momentum can be seen as a fat tail towards positive values
of $\epsilon$. M33 is a more model galaxy - it contains a well defined
rotationally supported thin disc as well as a central bulge and a
small, kinematically hotter, thick disc component.

While all this is certainly interesting in terms of galaxy formation,
the most important part to note -- with respect to this comparison
project -- is that all halo finders behave similarly when examining
this stellar decomposition.

For the decomposition of the gas particles we use a slightly different
method because some of the halo finders include the internal energy in
their unbinding procedure, while others do not. Unlike the stars, it
is thus unfair to decompose the gas distribution on energetic
grounds. Consider a co-rotating stellar and gaseous disc. Some halo
finders will use the internal energy of gas particles to unbind them
from this coherent motion and thus exclude them from the dynamical
decomposition described above. In this case, while all halo finders
will easily pick out star particles that are on circular orbits, not
all gas particles on circular orbits will be included. The results of
such a dynamical decomposition comparison are thus unreliable.

In order to overcome this inconsistency we use a method proposed by
\cite{Scannapieco10}. As with the stars, a $z$-direction is defined to
be the total angular momentum of all gas particles within 6, 8, and
10\hkpc\ for the MW, M31, and M33, respectively. We then calculate the
ratio of $\epsilon=J_{\rm z}/J_{\rm circ}$ where $J_{\rm circ}$ is the
angular momentum of a circular orbit at the same radius, i.e. $J_{\rm
circ} = r\times\sqrt{\frac{GM(<r)}{r}}$. In this way we bypass the
need to use the potential and kinetic energies since, as mentioned
earlier, gas particles also have thermal energies. When this ratio is
equal to unity, a gas particle is moving on a circular orbit (with
respect to the $z$-direction). Note that this ratio is in principle
unbounded from ($-\infty,\infty$). In practice however, few ($\sim1$~per cent
gas particles counter rotate and thus a lower limit of zero is
applicable.

The decomposition of the gas particles is shown in the bottom panels
of \Fig{fig:bdecomp-comp}. All gas distributions exhibit peaks at
$\epsilon=1$, indicating the presence of gaseous discs. M31 has a
wider distribution implying a thicker, dynamically hotter component
(as also suggested by the stellar decomposition above it). It is
interesting to note that in M31 \ahf\ and \rockstar\ identify a small
local minimum around $\epsilon=1$. This dip (with different maxima to
the left and the right) is interpreted as a subtle warping of the
gaseous disc. The astute observer will also notice a small but
non-negligible tail towards zero in the MW and M31 distributions;
these are gas particles whose angular momentum in the $z$-direction is
smaller than that of a circular orbit of this radius and who are thus
on either radial or inclined orbits.

The minor discrepancies between halo finders for the gas component are
certainly due to the differences in the handling of gas particles:
for instance, \ahf\ as well as \subfind\ read the gas thermal
energy taking it into account during the unbinding procedure; all
other codes do not consider the thermal energy at that stage (\stf\ and \rockstar) or
do not feature an unbinding procedure (\jumpd).

\section{Satellite Comparison} \label{sec:satellitecomparison}
We now turn to taking a closer look at the satellite populations of
the two prominent host haloes, i.e. all subsequent plots are based
upon a combined sample of MW and M31 subhaloes that reside inside a
sphere of radius 300 kpc centred on their respective host \textit{and} that
contain at least one star particle. Note that this radius is slightly larger that the virial radii of the two hosts
($\approx$ 200 kpc) and motivated by the observationally inferred value \citep{Watkins10}. We further impose a lower limit
for $V_{\rm max} \ge 10$km/sec as this corresponds to the completeness
level (cf. \Fig{fig:VmaxFunc} below). We list in \Tab{tab:Nsat} the
number of subhaloes compliant with these criteria found by each of the
participating finders.; and for \jumpd\ we give the number of baryonic objects found.
Further note that \jumpd\ is in the special situation of not considering subhalo finding and hence is not suited to be subjected to the common post-processing pipeline. Therefore, \jumpd\ only contributes to 
 the plots that merely require knowledge of the baryonic component. This data has been directly provided
by the finder itself. 

\begin{table}
  \caption{Number of subhaloes (compliant with our selection criterion, see text for details) found by each participating halo finder. For \jumpd\ we list the number of baryonic objects found.}
\label{tab:Nsat}
\begin{center}
\begin{tabular}{lccc}
\hline
 code 		& $N_{\rm sub}^{MW}$ 	& $N_{\rm sub}^{M31}$ 	& $N_{\rm sub}^{\rm combined}$\\
\hline
\ahf			&	32				&	43				&	75\\
\jumpd		&	25				&	35				&	60\\
\rockstar		&	29				&	38				&	67\\
\stf			&	31				&	42				&	73\\
\subfind		&	31				&	40				&	71\\
\hline
\end{tabular}
\end{center}
\end{table}

\subsection{\Vmax\ Function}\label{sec:VmaxFunc}
\begin{figure}
  \includegraphics[width=\columnwidth]{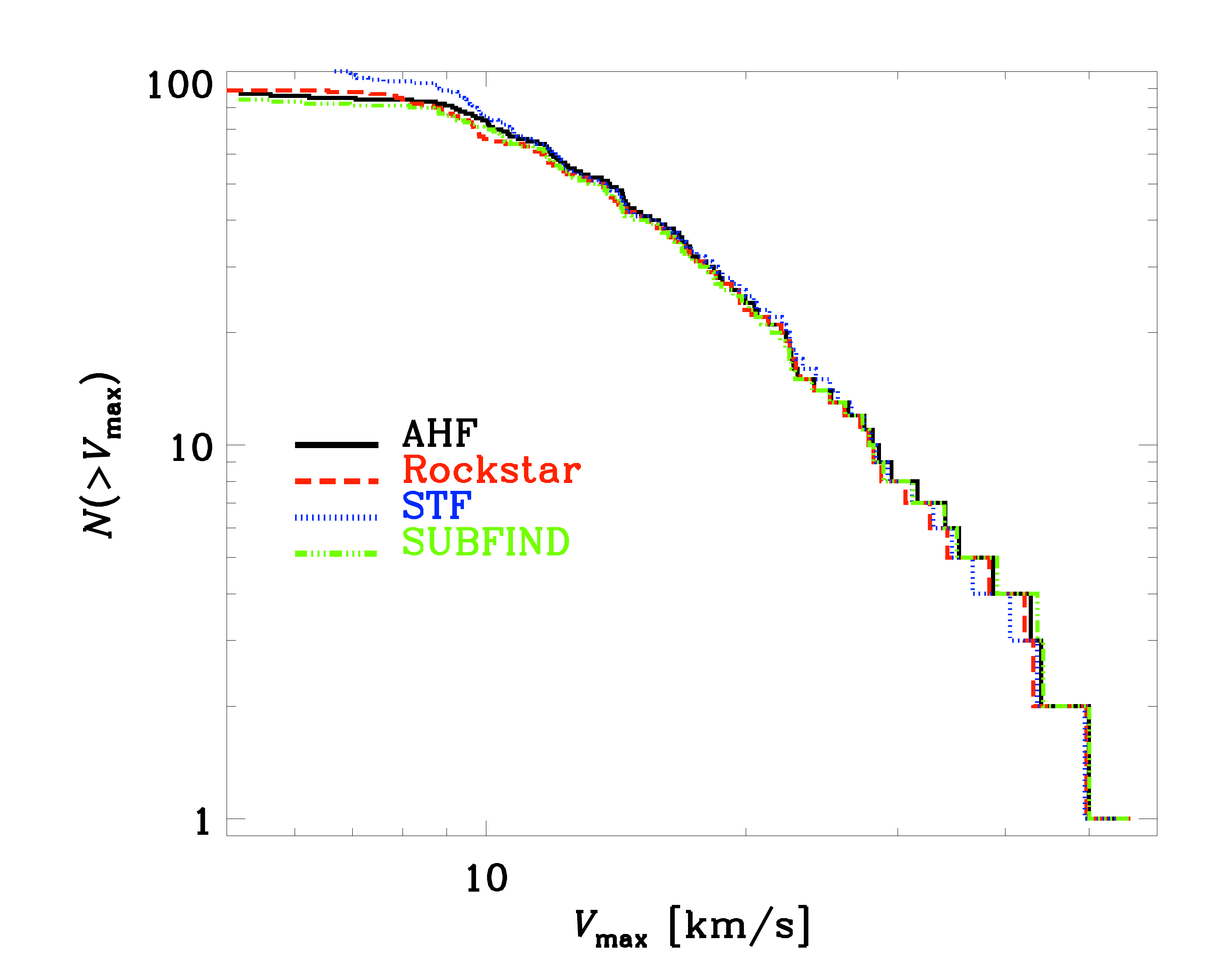}
  \caption{Cumulative \Vmax\ function for haloes closer than 300 kpc
  to either of the two simulated hosts M31 and MW that at least
  contain one star particle.}
\label{fig:VmaxFunc}
\end{figure}

We start with a quantity that has caught a lot of attention recently,
especially for the Local Group/Milky Way dwarf spheroidal galaxies
(dSph's): the peak of the circular velocity curve \Vmax\
\citep[e.g.][]{Boylan-Kolchin11,DiCintio11,Boylan11b,Vera-Ciro12,DiCintio12}. It
has, for instance, been shown by \citet{DiCintio11} that the inclusion
of baryonic physics into the simulation can lead to variations in
\Vmax\ (and \Rmax) when being compared to dark matter only
simulations; in particular, (sub-)haloes with high baryon fractions
tend to experience adiabatic contraction while objects with lower
baryon content move towards lower-\Vmax/higher-\Rmax\ values, possibly
due to mass outflows \citep{Navarro96b} or random bulk motion of gas
"heating" the central matter distribution \citep{Mashchenko06}.

In \Fig{fig:VmaxFunc} we compare the cumulative distribution of \Vmax\
found by all our finders. We find excellent agreement. Please note
that this is the only plot that has not been subjected to the lower
\Vmax\ limit. 

\subsection{\Rmax\ -- \Vmax\ relation}\label{sec:RmaxVsVmax}
\begin{figure}
  \includegraphics[width=\columnwidth]{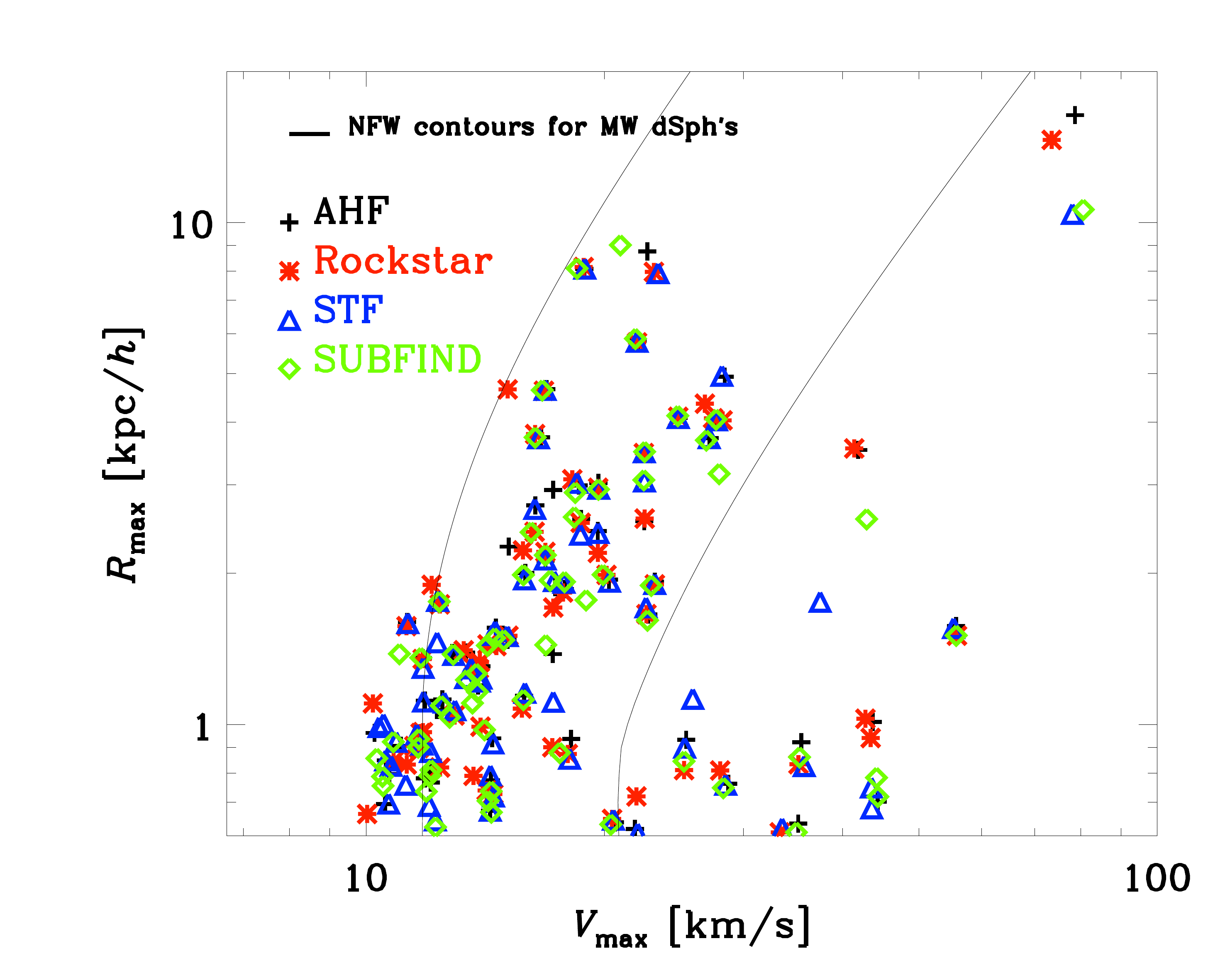}
  \caption{The relation between the peak of the circular velocity curve \Vmax\
  and its position \Rmax\ for subhaloes inside a 300 kpc sphere about
  either M31 or the MW. The thin solid lines delineate the $1\sigma$
  confidence interval of the observed bright MW dSph galaxies
  \citep[as in][]{Boylan-Kolchin11,DiCintio11}.}
\label{fig:RmaxVsVmax}
\end{figure}

As already outlined above when motivating the consideration of the
\Vmax\ function in the first place, there has been some substantial
debate about the failure of the most massive subhaloes found in
simulations of Milky Way type dark matter host haloes to host the most
luminous observed dSph's
\citep{Boylan-Kolchin11,DiCintio11,Boylan11b,Vera-Ciro12,DiCintio12}. To
verify whether different halo finders will contribute differently to
this discussion we present in \Fig{fig:RmaxVsVmax} the usual plot
confronting \Rmax\ with \Vmax. The two solid lines delimit the
$1\sigma$ confidence interval of the observed bright Milky Way dwarf
spheroidal galaxies, as defined in \citet[][please refer to these
references for more details]{Boylan-Kolchin11,DiCintio11}.

We again find remarkable agreement across the participating
codes, even though there appears to be some marginal scatter across
\Rmax\ values. This can be attributed to the fact that the
determination of \Vmax\ is certainly more stable for cosmological dark
matter haloes due to the flatness of the circular velocity curve; and this
flatness is obviously also responsible for (slight) variations in the
position of the peak velocity. In that regard, it should be noted that in
order to best calculate the position of the peak, i.e. \Rmax, in the
circular velocity curve we decided to only use the dark matter for this
purpose, i.e. \Rmax\ is defined as the position of the maximum of
$M_{\rm DM}(<r)/r$; however, $V_{\rm max}^2=GM_{\rm
tot}(<\Rmax)/\Rmax$ obviously takes into account all matter.

\subsection{Baryonic Mass}\label{sec:BaryonicMass}
\begin{figure}
  \includegraphics[width=\columnwidth]{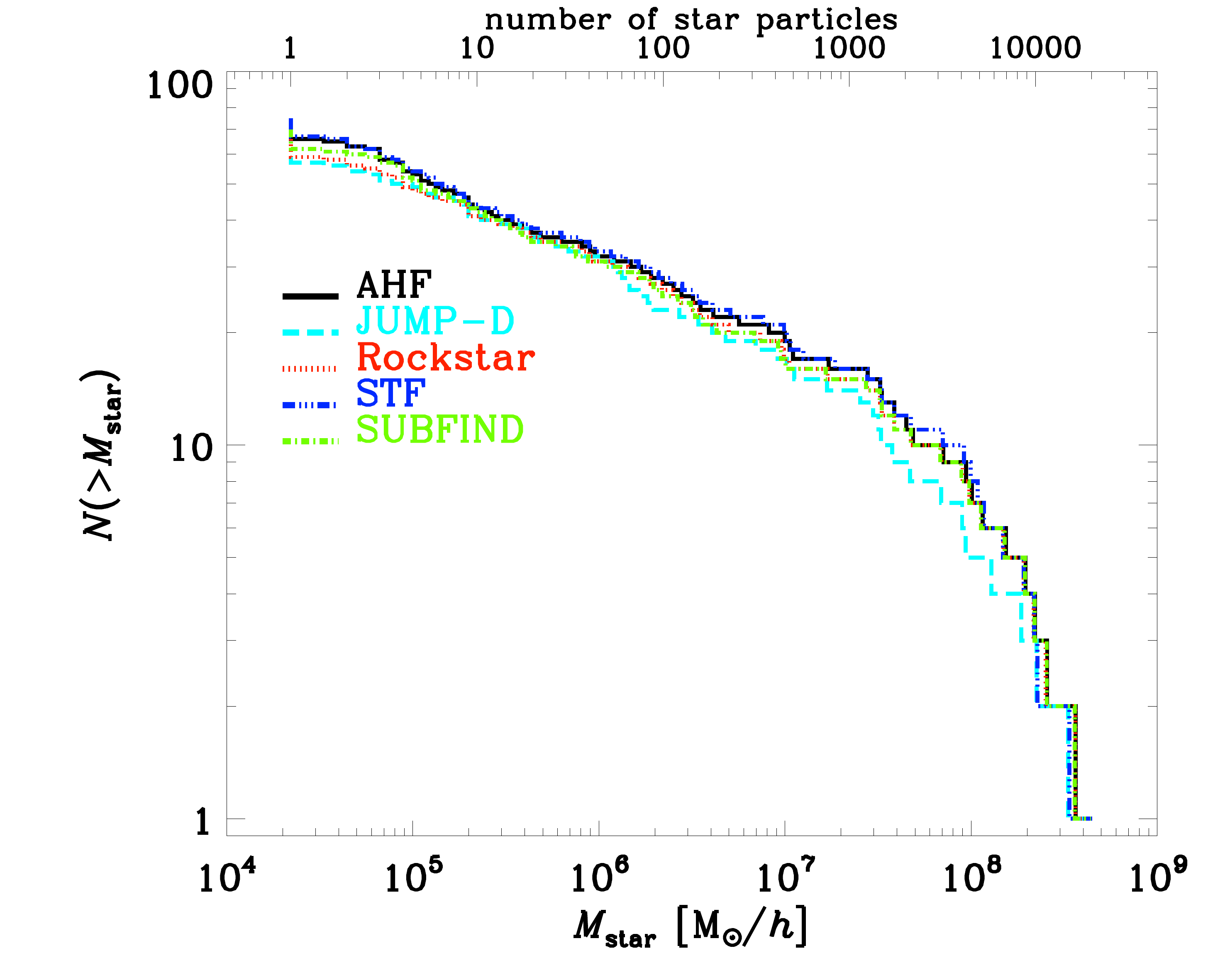}
  \includegraphics[width=\columnwidth]{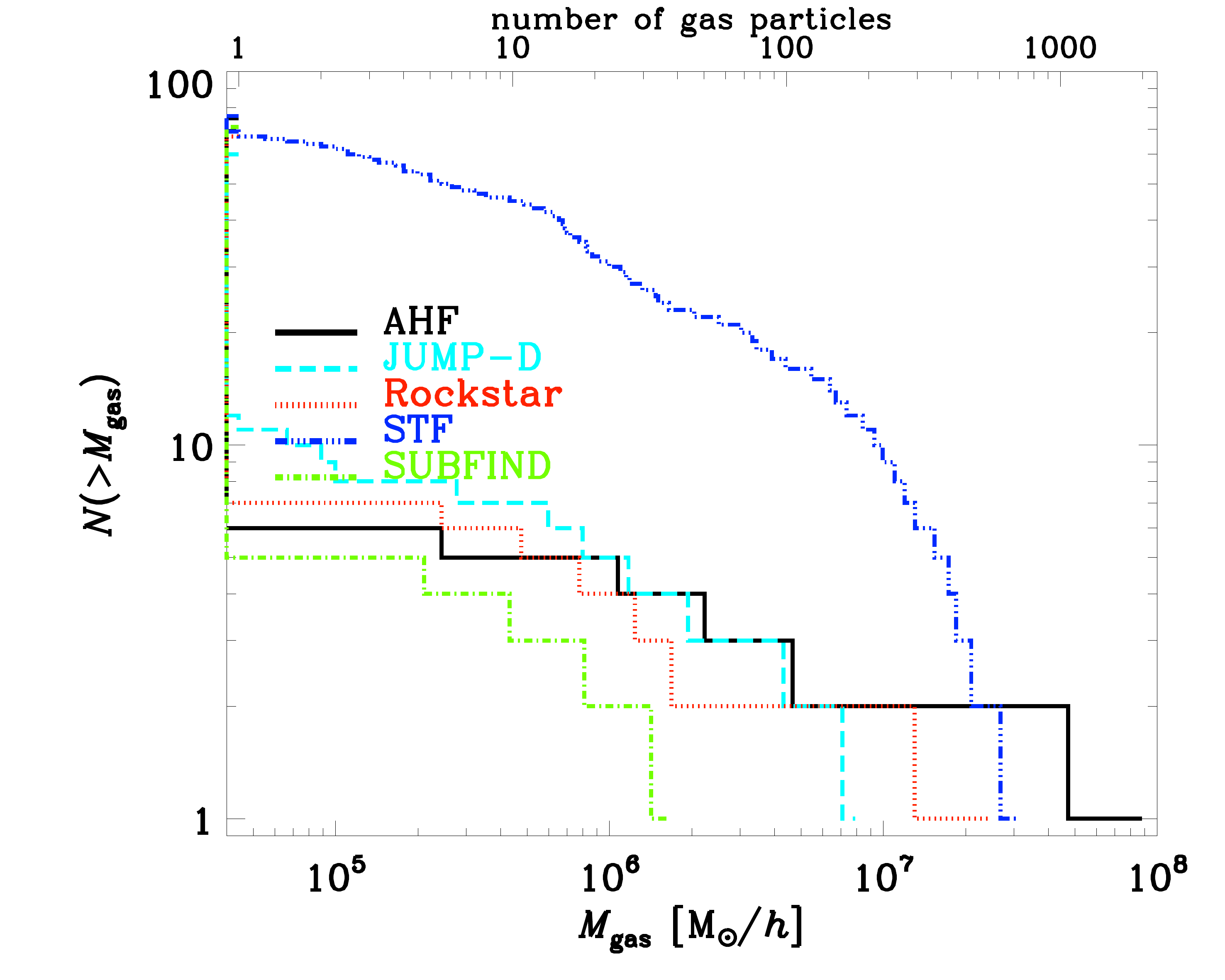}
  \caption{Cumulative stellar (top) and gas (bottom) mass functions
  for the same objects already shown in \Fig{fig:RmaxVsVmax}. The
  upper $x$-axis of each panel provides a translation of mass into
  number of particles.}
\label{fig:MstarGasFunc}
\end{figure}

In \Fig{fig:MstarGasFunc} we show the cumulative stellar (top) and gas
mass (bottom) function of all our subhaloes. The number of objects in the
stellar mass plot corresponds to the one given in \Tab{tab:Nsat}. But we also acknowledge that
this number is smaller than the total number of "subhaloes with stars" found
by the majority of finders as can be seen in \Fig{fig:VmaxFunc} due to the now applied \Vmax\ cut. 

While the
stellar masses of the objects agree remarkably well (note that \jumpd\ returns the stellar masses
of the objects within $r_{\rm star}$, and hence their slightly smaller values in some cases), we notice quite
substantial differences in the gas masses. While we return to the
differences found for the gas fractions in subhaloes in more detail
later, we bear in mind that the actual number of
subhaloes containing any gas is rather low for the majority of
finders (note that \jumpd\ returns the cold gas masses
within $r_{\rm gas}$ only when cold gas exist at the satellite center).
 \stf, which does not perform an unbinding procedure for the gas, finds a substantial number of satellites
containing gas particles down to the limit of one gas particle.  The
mass of a gas particle is $m_{\rm gas}=4.4\times 10^{4}$\hMsun\ which
corresponds to the lower limit of the $x$-axis.

\begin{figure}
  \includegraphics[width=\columnwidth]{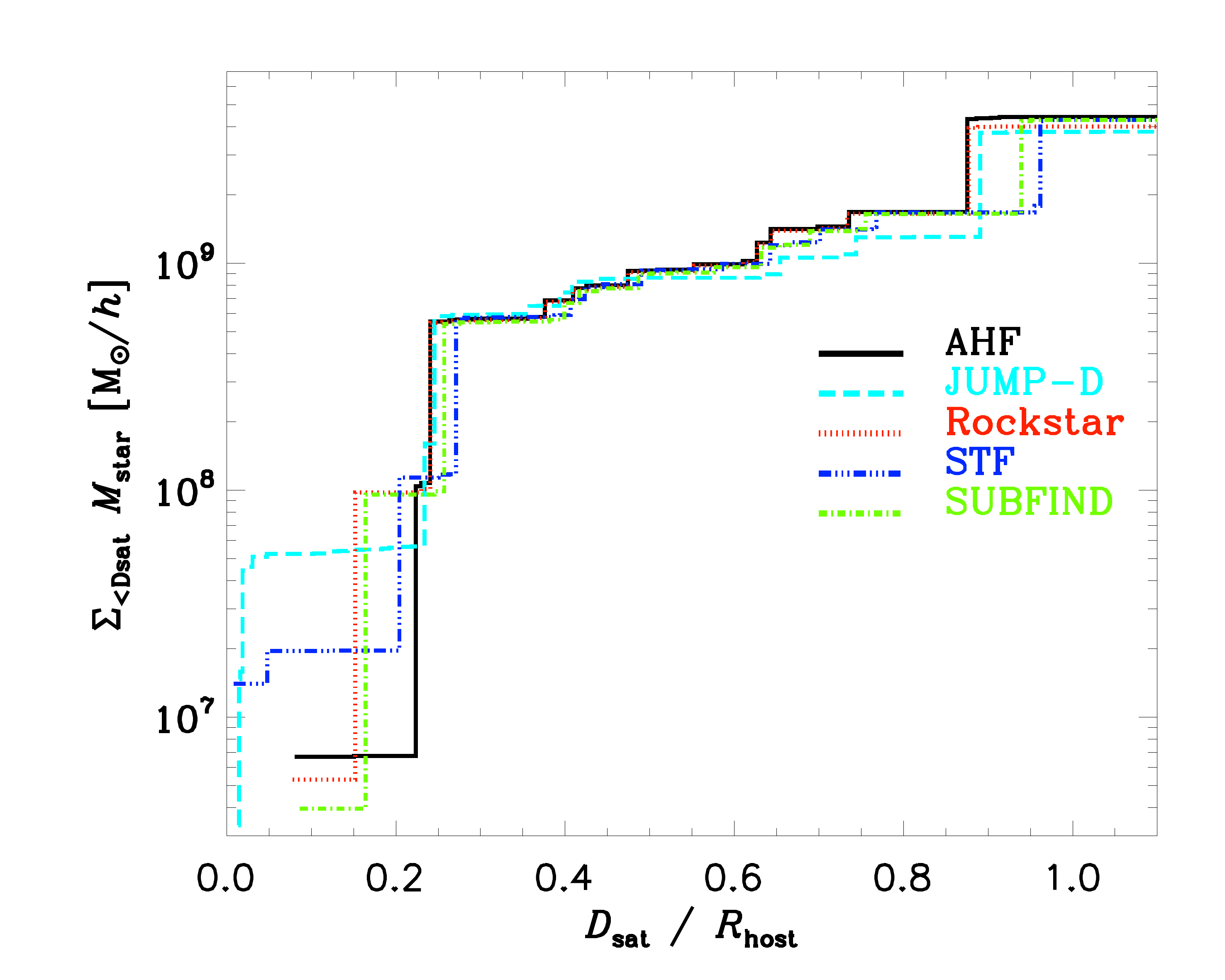}
  \caption{Cumulative stellar mass in objects as a function of distance to the (normalized) host centre.}
\label{fig:DistMstarFunc}
\end{figure}

While the agreement of the (stellar) masses appears to indicate that the objects found by the suite of finders used here are the same, this is not necessarily the case given their different modi of operations. To shed some more light on the cross-correlation of objects (in a statistical sense) we follow \citet{Onions12} and plot in \Fig{fig:DistMstarFunc} the cumulative stellar mass in objects as a function of distance from the centre of the host halo (normalized to the host radius). We stack the data from the MW and M31. The agreement proves that the satellite galaxies not only have similar masses but also live at approximately the same distance from the central galaxy. There are, however, a few (massive) objects found by \jumpd\ and \stf\ close to the centre not picked up by the other finders. There is another interesting feature in this plot, namely that \jumpd\ is slightly below the other finders at larger radius, which is possibly due to ignoring the underlying dark matter, so that either the individual galaxies have less stars or some with a low stellar content are missed because without the dark matter particles they are too small to be assigned as reliable substructures.

\begin{figure}
  \includegraphics[width=\columnwidth]{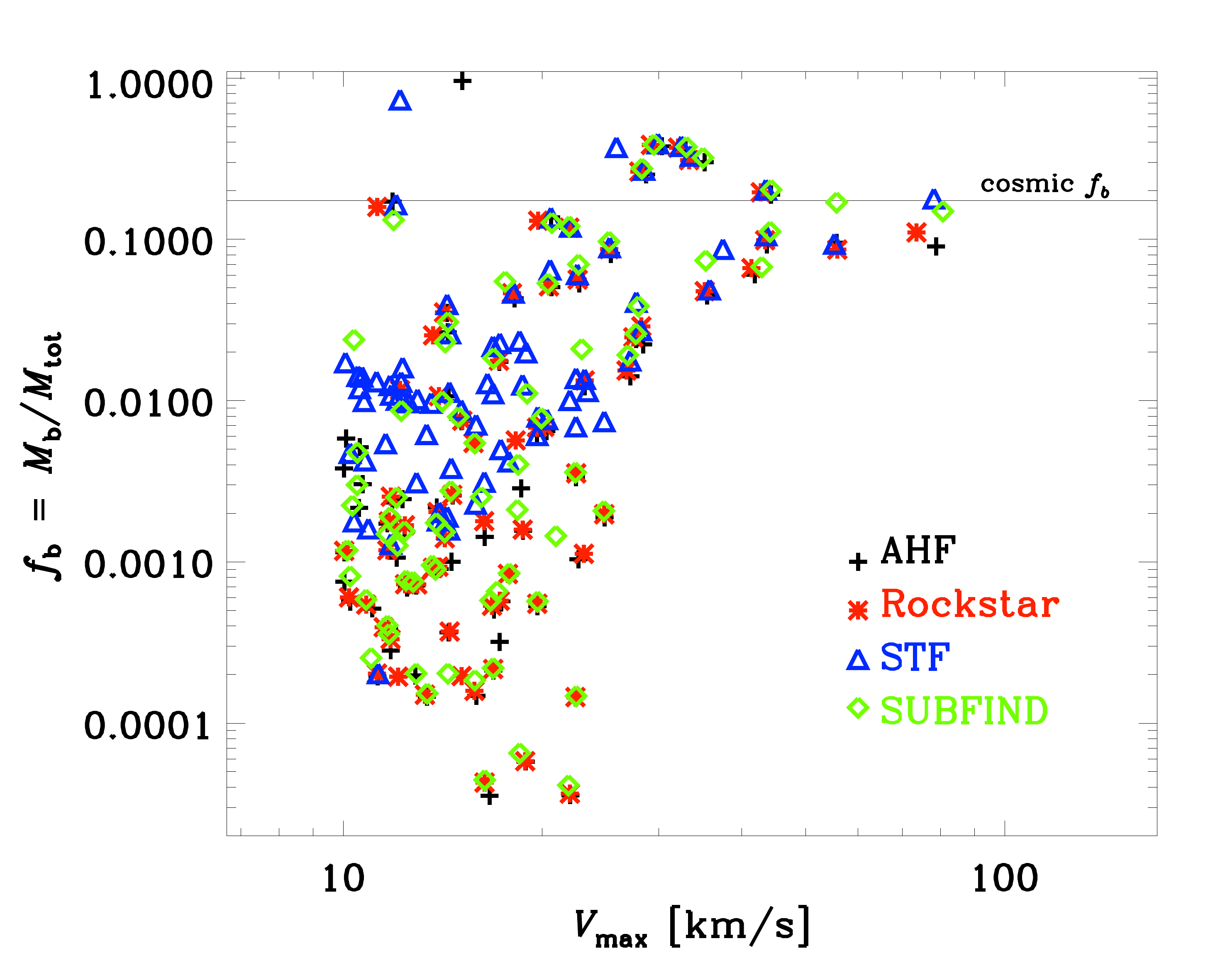}
  \caption{Baryon fraction $f_b$ vs. \Vmax.}
\label{fig:fbVSvmax}
\end{figure}

So far we have separated stellar and gas mass when studying
baryons. But we now briefly turn to the baryon fraction, investigating
its relation with \Vmax\ in \Fig{fig:fbVSvmax}. Again, all codes show
the same tendancy for more massive subhaloes to also contain more
baryons and it is difficult to decipher any differences (despite our
knowledge that there are marked differences in the gas content),
unless we move to a comparison of objects on a one-to-one basis to be
undertaken in \Sec{sec:crosscomparison}. However, we note that \stf\
does not find as many low-$f_b$ objects (for $\Vmax<20$ km/sec) as the other finders due to
the fact that it keeps more gas in its subhaloes as seen in (the
bottom panel of) \Fig{fig:MstarGasFunc}. Bear in mind again that
\stf\ did not pass the gas particles through an unbinding routine,
hence the expected large gas fractions at all mass scales.

\subsection{Luminosity Function}\label{sec:MVfunc}
\begin{figure}
  \includegraphics[width=\columnwidth]{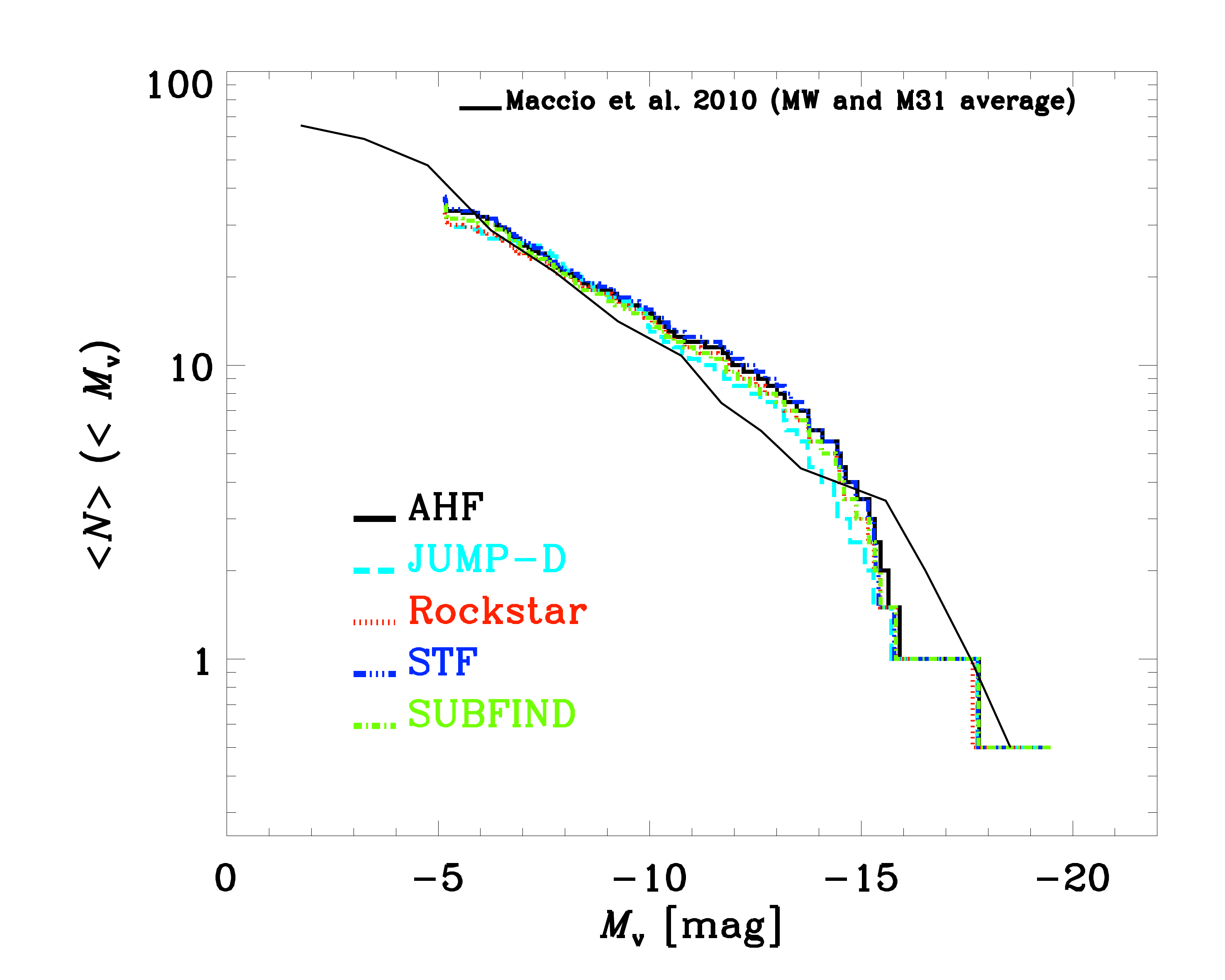}
  \caption{The luminosity function of subhaloes in the Johnson
  V-Band. The ``Maccio'' observational data (thin solid line) is a
  combination of the volume corrected MW luminosity function
  \citet{Koposov08} augmented with information from \citet{Mateo98}
  and \citet{Maccio10} under the assumption of an NFW-like radial
  distributions of satellites. Note that the comparison to the
  observational data is \textit{not} the prime target of this study
  and only serves as a reference.}
\label{fig:MVfunc}
\end{figure}

In \Fig{fig:MVfunc} we now present the Johnson V-Band luminosity
function as well as the observational data as taken from
\citet{Koposov08} and \citet{Maccio10}, respectively (thin solid line,
referred to as ``Maccio sample''): these data are a combination of the
volume corrected MW satellite luminosity function \citep{Koposov08}
augmented with information from \citet{Mateo98} and \citet{Maccio10}
kindly provided to us by Andrea Maccio (personal communication). And
even though all luminosity functions agree with the Maccio sample
rather well, we stress that we included the observational data merely
as a reference to guide the eye. It is not our prime objective to
reproduce the MW and/or M31 luminosity function of satellite galaxies
with our simulation. We find again excellent agreement between the
finders. But his comes at no surprise as we found negligible
variations in the stellar mass of subhaloes in (the upper panel of)
\Fig{fig:MstarGasFunc} already.

\section{Satellite Cross-Comparison} \label{sec:crosscomparison}
While the general agreement for the satellite populations between codes is rather outstanding -- at least when it come to the fundamental characteristics such as \Vmax\ and luminosity $M_V$ -- we now go one step further and directly compare properties of objects on an individual basis. To this extent we use the same matching criterion (based upon the uniqueness of particle IDs) successfully used to construct merger trees or cross-correlate different simulations before \citep[e.g.][]{Klimentowski10,Libeskind10,Knebe11a} by facilitating a tool that comes with the \ahf\ package called \texttt{MergerTree}. Even though originally designed to identify corresponding objects in the same simulation at different redshifts it can equally be applied to find "sister objects" in an analysis of the same snapshot done with a different halo finder.

Note that distribution functions (as well as scatter plots) previously presented will not only suffer from differences in individual halo properties but also encode the fact that some finders may have identified different numbers of objects (cf. \Tab{tab:Nsat}). To circumvent this we aim at directly comparing quantities on a halo-to-halo basis and move from general distribution functions and their variations as discussed above to cross-comparing satellites. The plots in the following Sub-Sections~\ref{sec:errorVmaxRmax} through to \ref{sec:errorMV} now all follow the same logic: the $x$-axis shows the median of the haloes' \Vmax, whereas the $y$-axis gives the normalised difference between the lower and upper percentiles equivalent to the 2nd and 3rd ranked of the distribution across all halo finders (\jumpd\ is not included in these plots as no dark matter halo required for the calculation of \Vmax\ has been provided). We deliberately chose to use medians and percentiles as the distribution of properties across finders might be non-Gaussian and at times biased by just one outlier. Further note that not all subhaloes have an error greater than zero: those for which the difference in the 3rd and 2nd ranked finder is zero have been artificially placed at the bottom of each panel at a value of $1.2 \times 10^{-4}$ yet entered with their correct value of zero into the calculation of the error percentages given below in the respective Sub-Section. We further indicate the 1 per cent and 10 per cent error by a horizontal dashed line in each plot and provide the fraction of those objects below that error as a legend.

\subsection{A Common Set}\label{sec:commonset}
For a given subhalo in one catalogue we locate the subhalo which
contains both the greatest fraction of its particles and which is
closest to it in mass. Specifically, for each sister candidate $i$ we
calculate a merit $m_{i}$, defined as

\begin{equation}
  m_{i}=\frac{n^{2}_{\rm shared,i}}{n_{\rm reference} \ \ n_{\rm sister, \it i}}
\label{merit}
\end{equation}

\noindent
where $n_{\rm shared,i}$ is the number of particles shared between
both subhaloes, $n_{\rm sister,\it i}$ is the total number of
particles in the sister candidate and $n_{\rm reference}$ is the total
number of particles in the reference subhalo. We then identify the
sister as that halo with the largest value of $m_{i}$. Typically,
successful matches result in $m_{i} \gsim 0.5$ and we assume to have
failed to find a sister if $m_{i} < 0.2$ for a given subhalo.\footnote{We found these criteria
to give the most reliable merger trees for subhaloes \citep[e.g.][]{Klimentowski10,Libeskind10infall,Knebe11b,DiCintio11,Libeskind11,DiCintio12} and hence decided to apply
them also for the cross-correlation.}

By restricting ourselves to the set of objects found by \textit{every}
halo finder we are able to directly compare the properties of
\textit{the same object} across all finders. Even though \ahf\ found
the most haloes compliant with our initial selection criterion and we
chose to use its catalogue as the basis, we confirm that every subhalo
found by \ahf\ has an actual counterpart in all other
analyses. However, some of these counterparts do not fully lie within
the required 300 kpc-sphere and hence were rejected from the previous
analysis. Using them now will increase our statistics without
obscuring the results: the aim is to compare the same objects across
different finders and they certainly serve this purpose.

\subsection{\Vmax\ \& \Rmax}\label{sec:errorVmaxRmax}
\begin{figure}
  \includegraphics[width=\columnwidth]{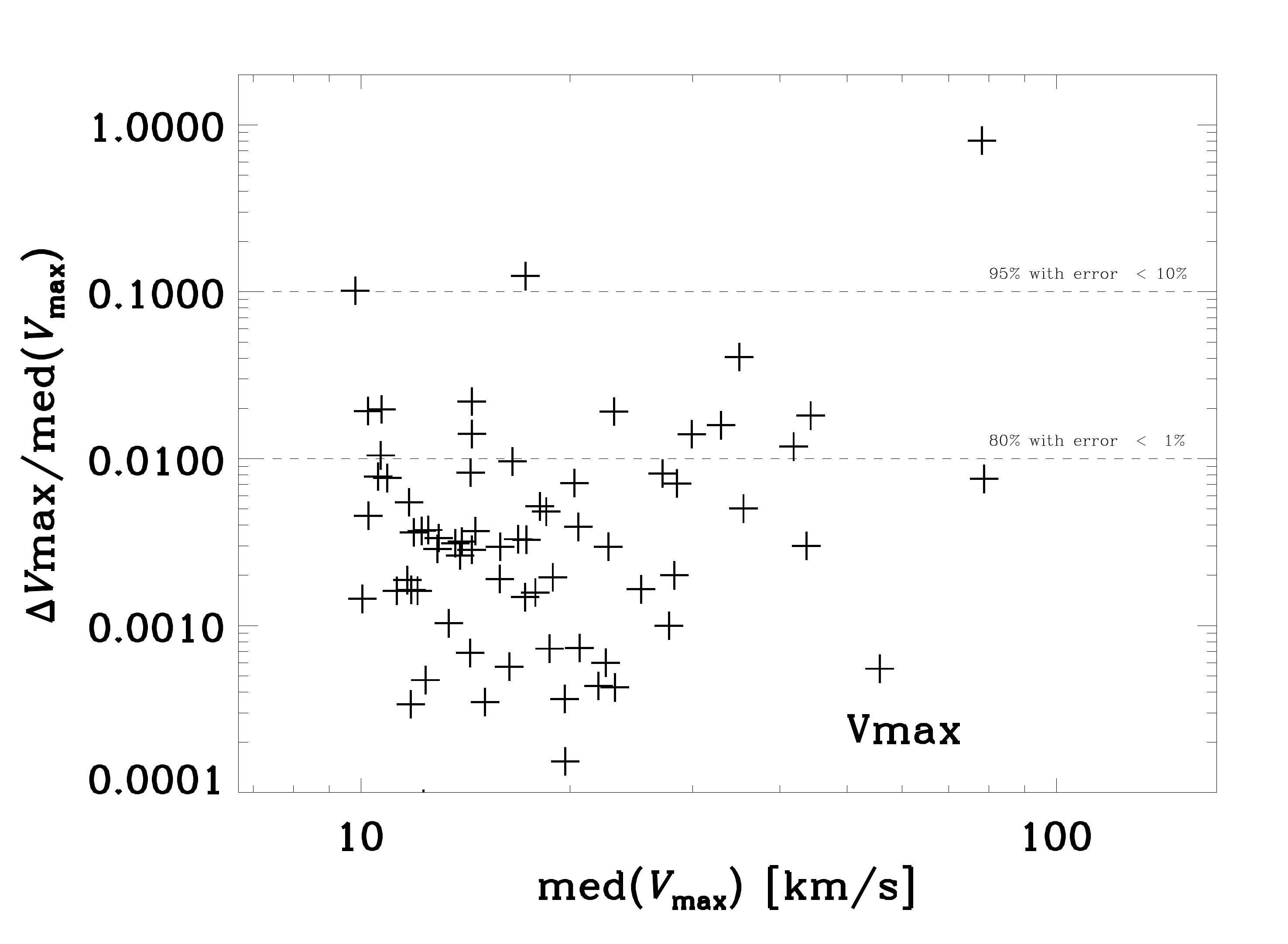}
  \includegraphics[width=\columnwidth]{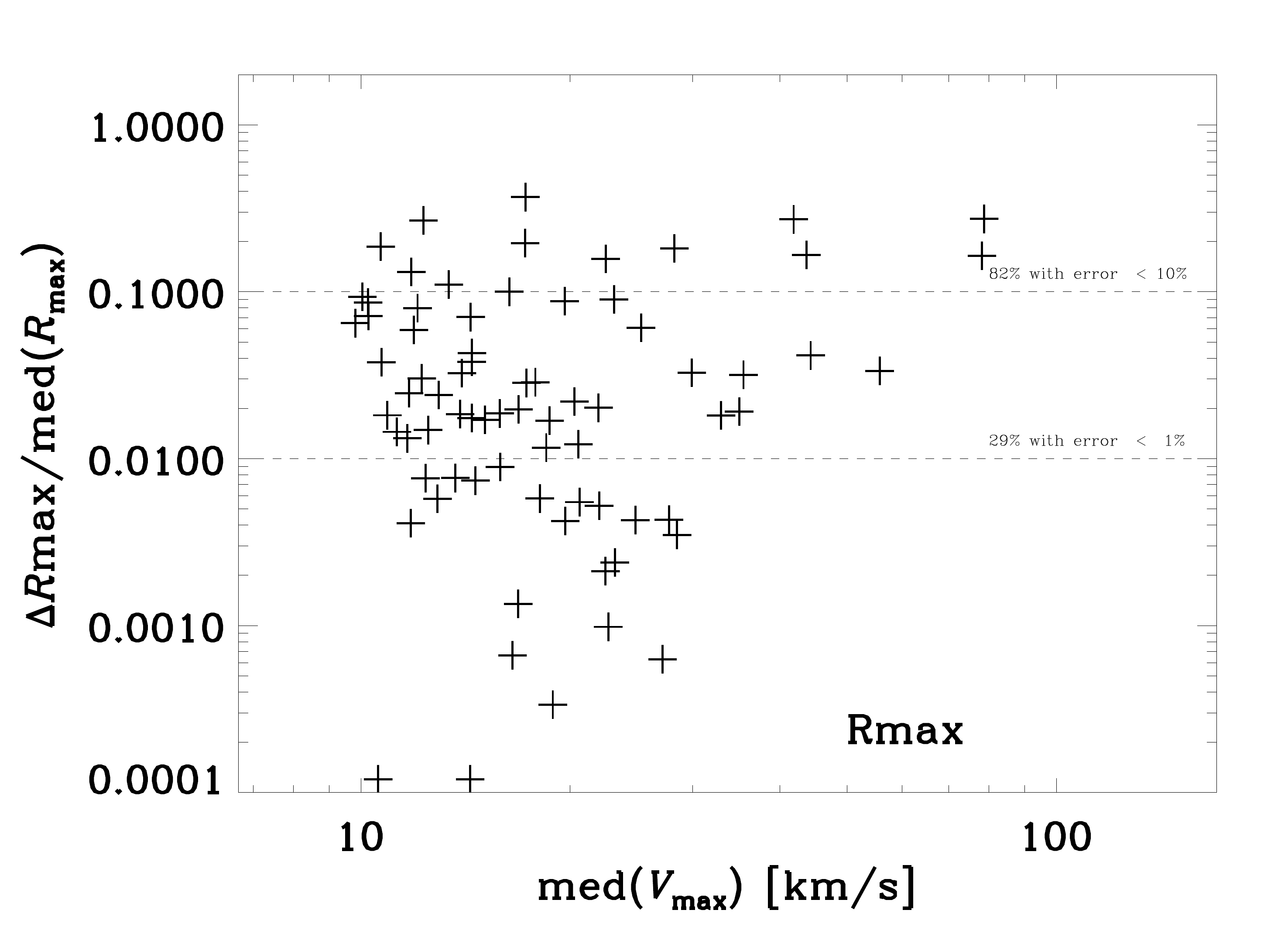}
  \caption{Relative error in \Vmax\ (top) and \Rmax\ (bottom). The
  horizontal dashed line simply indicates an error of 1 per cent for
  reference. Subhaloes appearing at an error value of $1.2\times
  10^{-4}$ are the ones that actually have zero values.}
\label{fig:errorVmaxRmax}
\end{figure}

While we have seen that the general relation between the codes for
subhaloes compliant with our selection criterion is rather remarkable,
\Fig{fig:errorVmaxRmax} further confirms that also the one-to-one
scatter of cross-identified objects is small: inspecting the
differences in the peak of the circular velocity curve \Vmax\ and its location
\Rmax, we find the overall median values of the error to be $<1$~per cent and
$2$~per cent for \Vmax\ and \Rmax, respectively. It is evident that \Vmax\
gives better agreement as it is a more stable property than \Rmax\ due
to the flatness of the circular velocity which is responsible for at
least a part of the scatter in \Rmax.

\subsection{Baryons}\label{sec:errorBaryons}
\begin{figure}
  \includegraphics[width=\columnwidth]{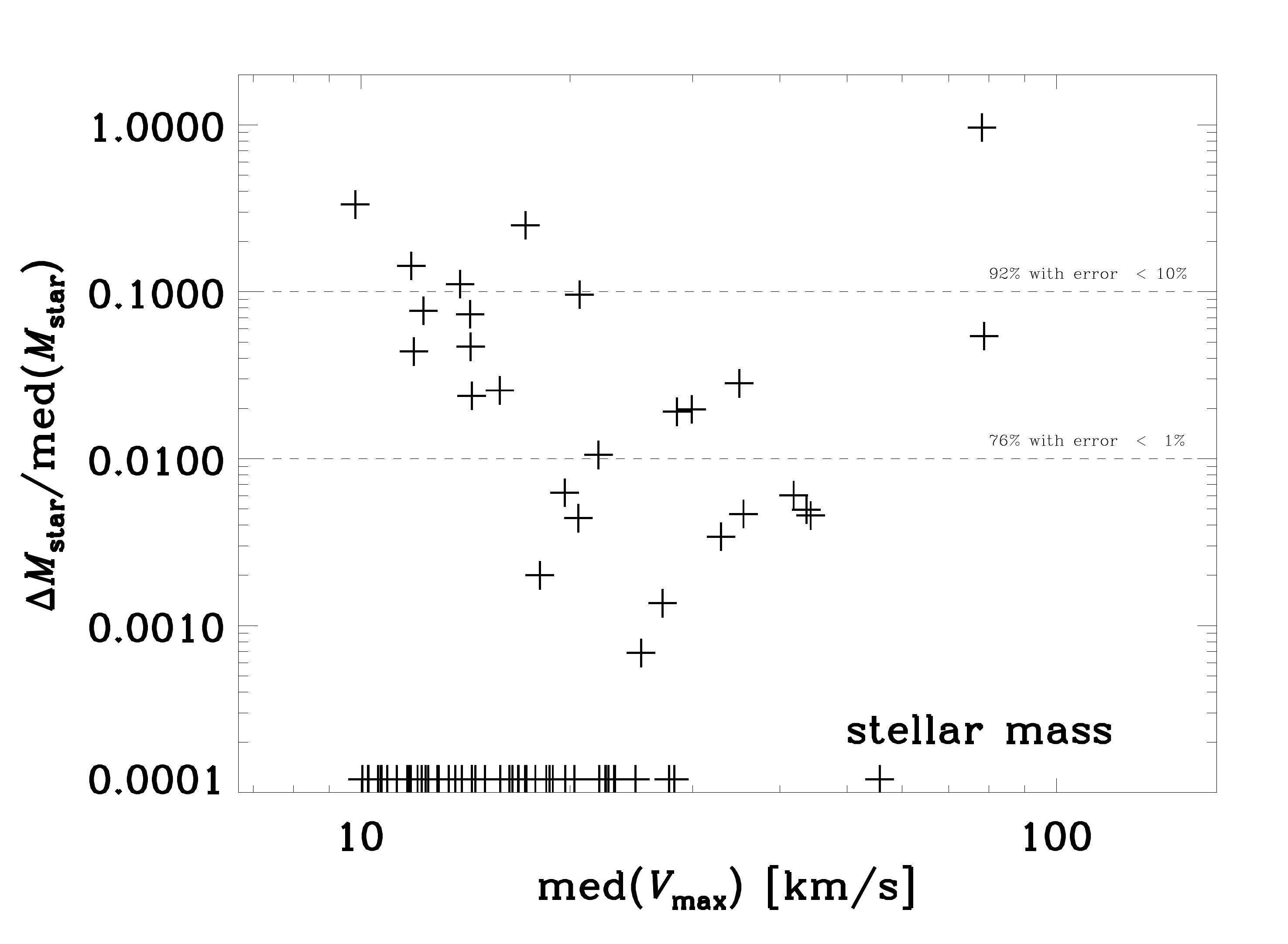}
  \includegraphics[width=\columnwidth]{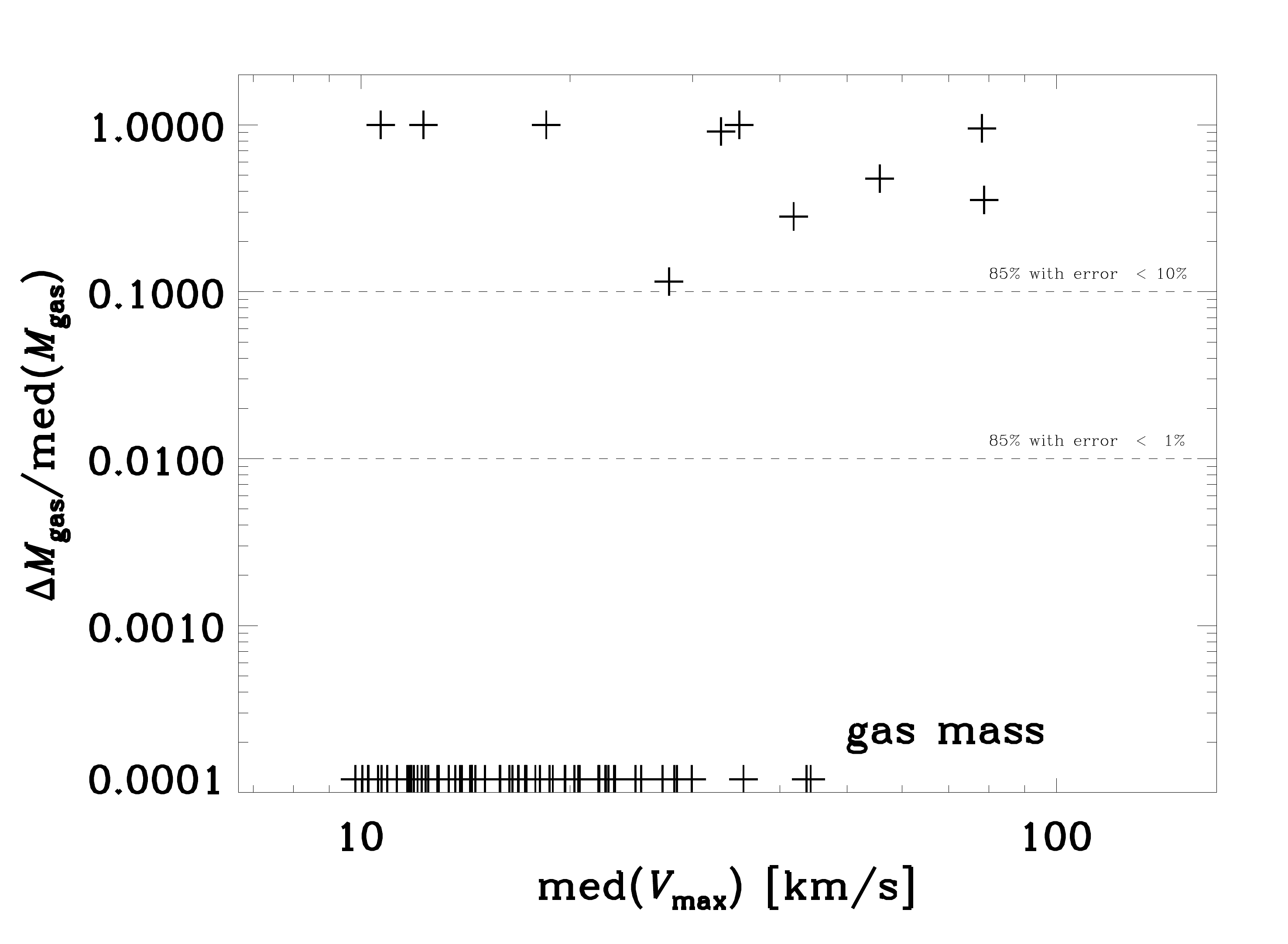}
  \caption{Same as \Fig{fig:errorVmaxRmax} but for stellar mass $M_*$
  (top) and gas mass $M_{\rm gas}$ (bottom).}
\label{fig:errorMstarMgas}
\end{figure}

The comparison of the gas mass function in \Fig{fig:MstarGasFunc}
already revealed differences while the stellar mass
function showed excellent agreement. It therefore comes as
no surprise that we find a median error of $86$~per cent for the gas mass
$M_{\rm gas}$ (considering only subhaloes that actually contain gas)
and $<1$~per cent for the stellar mass $M_*$ in \Fig{fig:errorMstarMgas}. As
already noted, some subhaloes may be missing from the plot if the
difference between their 3rd and 2nd ranked finders is zero. This is
in fact the case for some of the objects entering the stellar and gas
mass comparison: 47 (out of the 75 successfully cross-matched
subhaloes) show identical stellar masses and 63 have no gas particles
at all. In that regard, the actual median error of the gas mass is
$0$~per cent (while the stellar mass value remains unaffected), but we
decided to rather report the deviation only considering subhaloes
containing gas and seen in the plot, respectively.

\begin{figure}
  \includegraphics[width=\columnwidth]{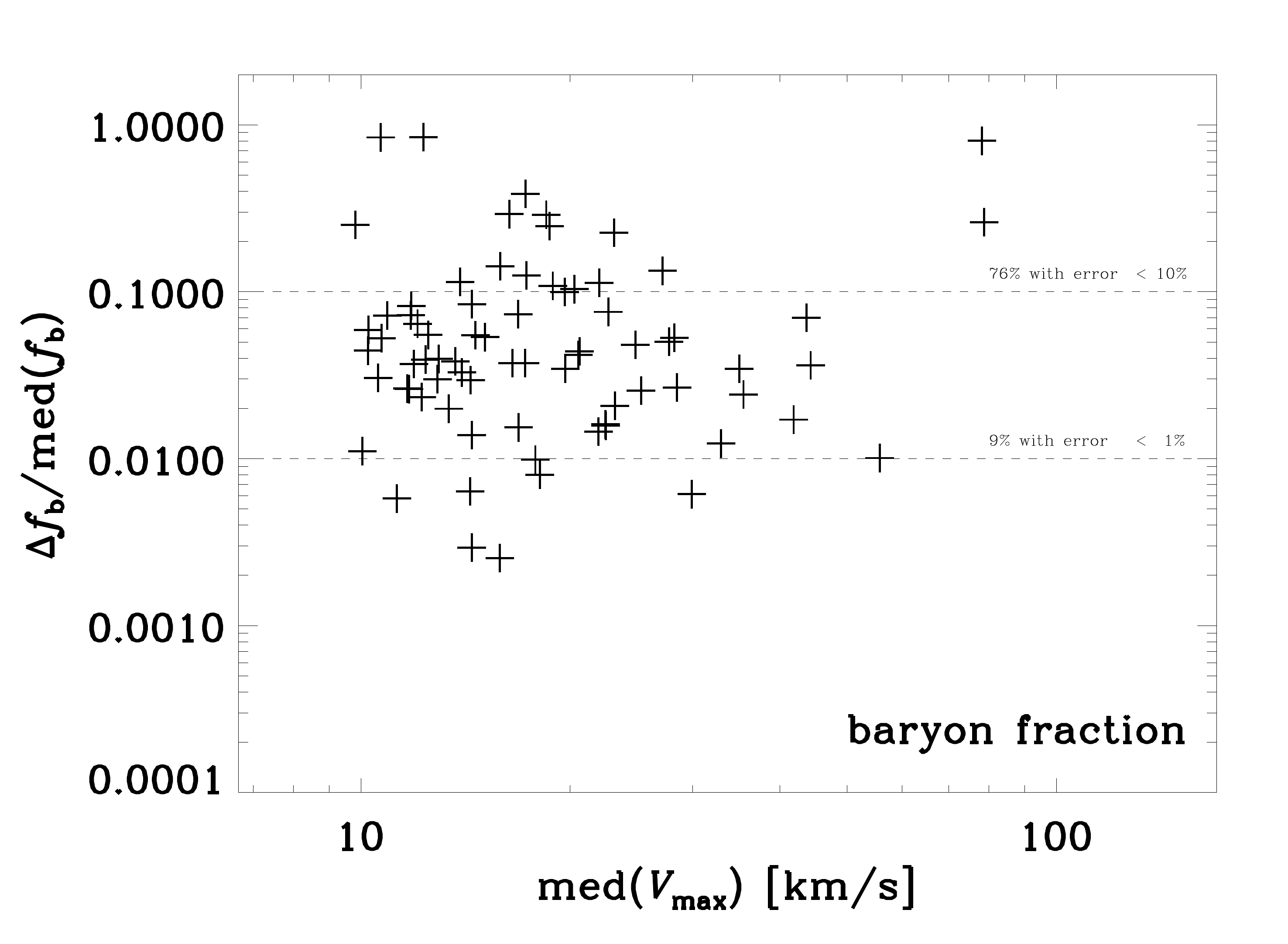}
  \caption{Same as \Fig{fig:errorVmaxRmax} but for the baryon fraction
  $f_b=(M_*+M_{\rm gas})/M_{\rm tot}$.}
\label{fig:errorfb}
\end{figure}

How does the error in baryons with the simultaneous agreement in
stellar mass translate into differences in the baryon fraction of
subhaloes? This can be verified in \Fig{fig:errorfb} where we find an
overall scatter of $5$~per cent. This leads to the conclusion that the
stellar mass content is by far the most dominant and important for the
baryon fraction and the substantial differences in gas mass only
affect its value marginally.

\subsection{Luminosity}\label{sec:errorMV}
\begin{figure}
  \includegraphics[width=\columnwidth]{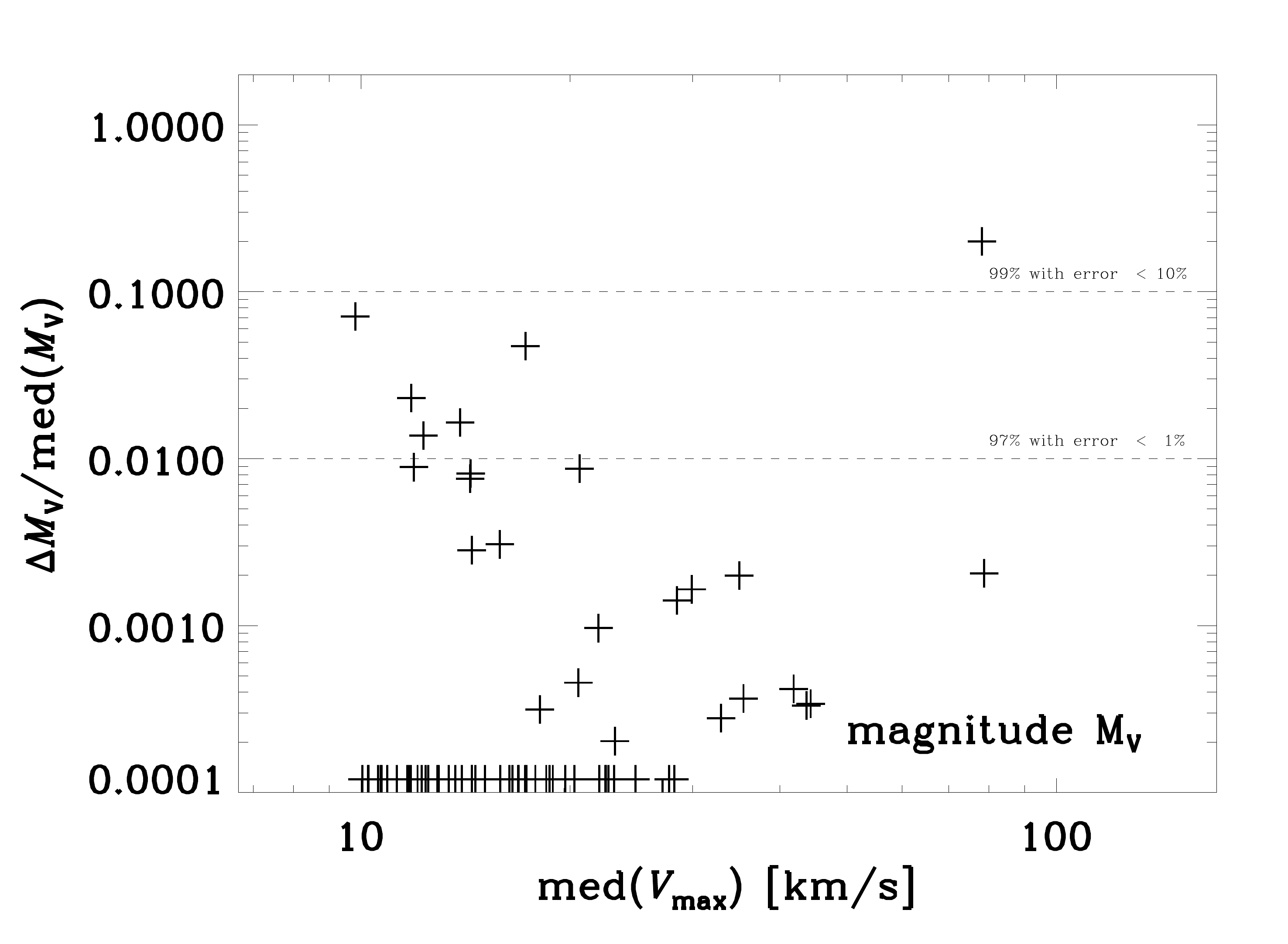}
  \caption{Same as \Fig{fig:errorVmaxRmax} but for the Johnson V-band
  magnitude $M_V$.}
\label{fig:errorMV}
\end{figure}

While the agreement between codes for the V-band magnitude was rather
excellent -- as verified by the luminosity function presented in
\Fig{fig:MVfunc} -- will we find similar consistency when comparing
objects on an individual basis? This can be verified in
\Fig{fig:errorMV} where we find a general median error of $<1$~per cent
(noting again that 47 out of 75 cross-matched subhaloes contain an
equal amount of star particles). One thing to note here is that the
errors for $M_V$ are a fair bit smaller than the differences found for
the stellar masses, even though luminosities are directly related to
the star particles (and their properties like age and
metallically). But while the transformation from the upper panel of
\Fig{fig:errorMstarMgas} to \Fig{fig:errorMV} is most certainly
non-linear, the general features (and "satellite placement" in the
plot) are nevertheless preserved; the objects are only shifted towards
smaller errors entailing that luminosity as measured here via $M_V$ is
not as sensitive to the stellar content as the mass itself.

\subsection{Mass Distributions}\label{sec:DensProfileSubs}
\begin{figure*}
  \includegraphics[width=2.15\columnwidth]{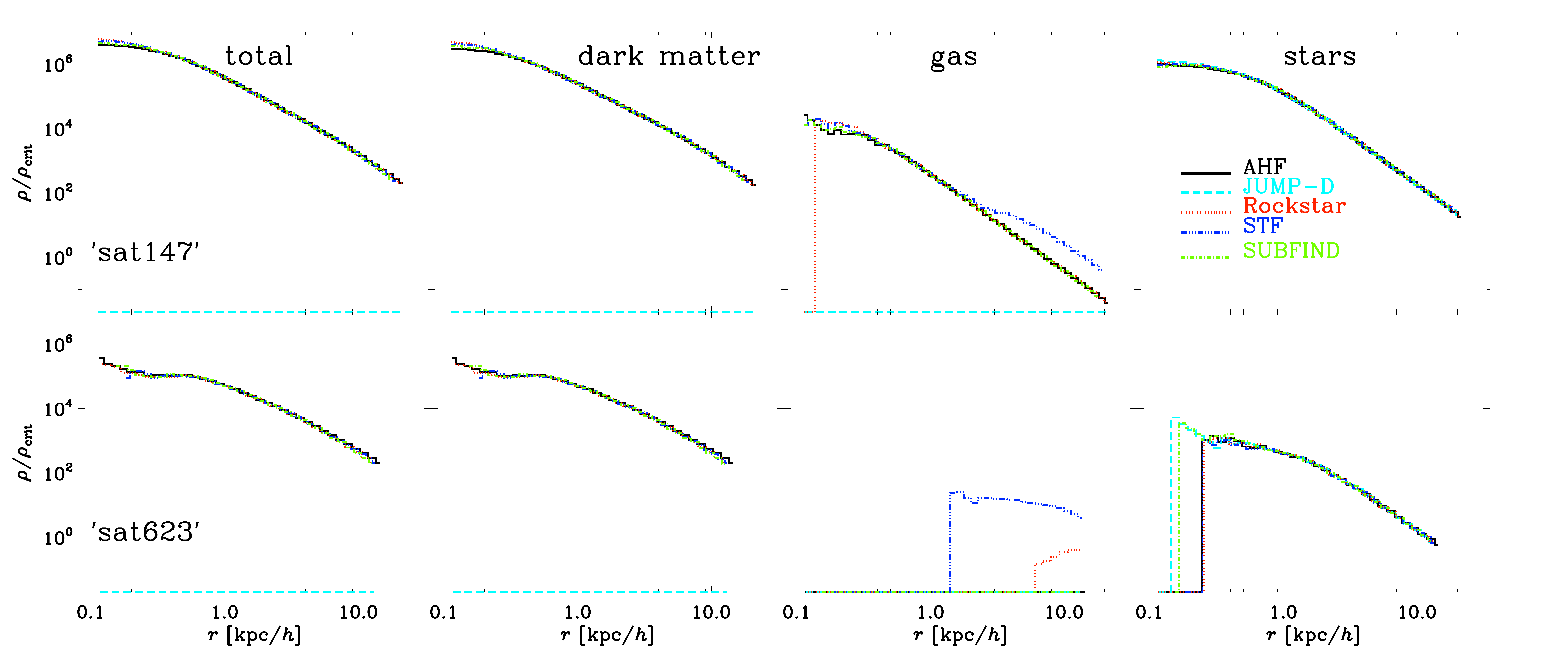}
  \caption{Density profile of the total (first column), dark (second
  column), gas (third column), and stellar (fourth column) matter for
  a massive (upper row, called 'sat147') and a low-mass (lower row,
  called 'sat623') subhalo.}
\label{fig:DensProfileSubs}
\end{figure*}

\begin{figure}
  \includegraphics[width=\columnwidth]{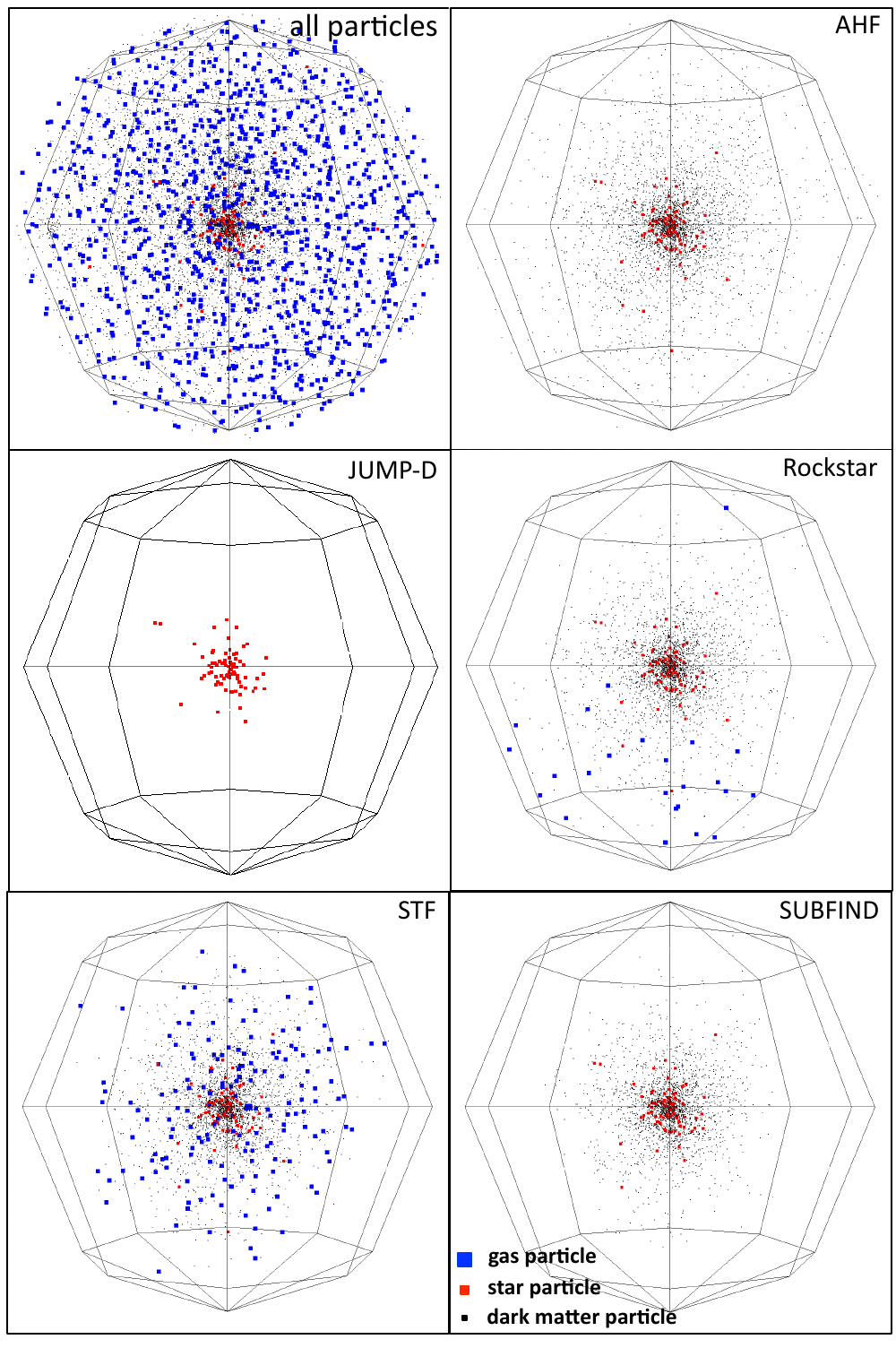}
  \caption{Visualisation of a subhalo showing all particles inside a
  spherical region about the identified centre (upper left panel) vs. the
  actually identified particles of the individual halo finder showing all types of particles: gas (blue), stars (red), and
  dark matter (black). Note that the object coincides with the lower
  panel object of \Fig{fig:DensProfileSubs}.}
\label{fig:ParticleComparison}
\end{figure}

To further explore the differences -- in particular for the gas
content of subhaloes -- we look at the mass distribution of the
various components (i.e. dark matter, gas, and stars, as already done
for \Fig{fig:DensProfile}) in the subhaloes found by \ahf\ and
cross-matched to all other halo finders. In \Fig{fig:DensProfileSubs}
we showcase two examples, one massive subhalo (top row, containing
roughly 20000 particles) and one less massive object (bottom row,
containing 3500 particles); note that using objects with even lower
numbers of particles will no longer provide credible profiles.
From left to right the columns show the total matter profile, the dark matter, the gas, and the stellar mass profile (all in units
of the critical density at redshift $z=0$). While we can again confirm
rather excellent agreement for the dark matter and stars, we clearly
see differences in the gas -- as expected from the previous plots. In
particular, the lower mass subhalo does not feature a gas profile at
all for the majority of the applied halo finders. Note again that all
curves terminate at the edge of the object.

\subsection{Gas Treatment -- A Test Case}\label{sec:GasTreatment}
To better understand the differences seen in the gas properties in
Figs.~\ref{fig:MstarGasFunc}, \ref{fig:errorMstarMgas} as well as
\ref{fig:DensProfileSubs} we went back to the simulation data and
extracted all the particles in a spherical region about the centre of
the subhalo featured in the lower panel of
\Fig{fig:DensProfileSubs}. The results can be viewed in \Fig{fig:ParticleComparison}. We can clearly see that the region about
the object's centre contains a substantial number of gas particles
(shown in the upper left panel). But all codes featuring a treatment of the
gas thermal energy either during or prior to the unbinding (i.e. \ahf\ and \subfind) remove essentially all gas from
the subhalo; the other two finders that do not include the thermal
energy during the unbinding are left with a residual amount of gas.

When using \ahf\ in the case where the gas thermal energy has been
ignored, \ahf\ basically considers \emph{all} gas particles seen in
the left panel to be part of the subhalo. In contrast, due to their
phase-/velocity-space nature, both \rockstar\ and \stf\ consider the
majority of the gas particles to belong to the background host and
keep only a small amount of them. For the object considered here the
effective thermal velocity of each gas particle is always larger than
its kinetic velocity (not shown here) and hence the grouping in
phase- or velocity-space will naturally remove (hot) gas whenever a
gas particle is considered not belonging to it based upon kinetic
velocity only. Or put differently, the gas component forming part
of the background halo is prone to be removed by such finders as they
inherently use velocity information when grouping and collecting the
initial set of particles, whereas configuration space finders only
deal with velocities (either kinetic or thermal) in a
(post-processing) unbinding procedure. On a side note, a visual
inspection of a larger region about this particular sample satellite
galaxy indicates that it has passed extremely close to its host
already and been subjected to severe tidal forces; this might also
explain why \rockstar\ appears to only have found one side of the
object.

\begin{figure}
  \includegraphics[width=\columnwidth,height = 0.45\columnwidth]{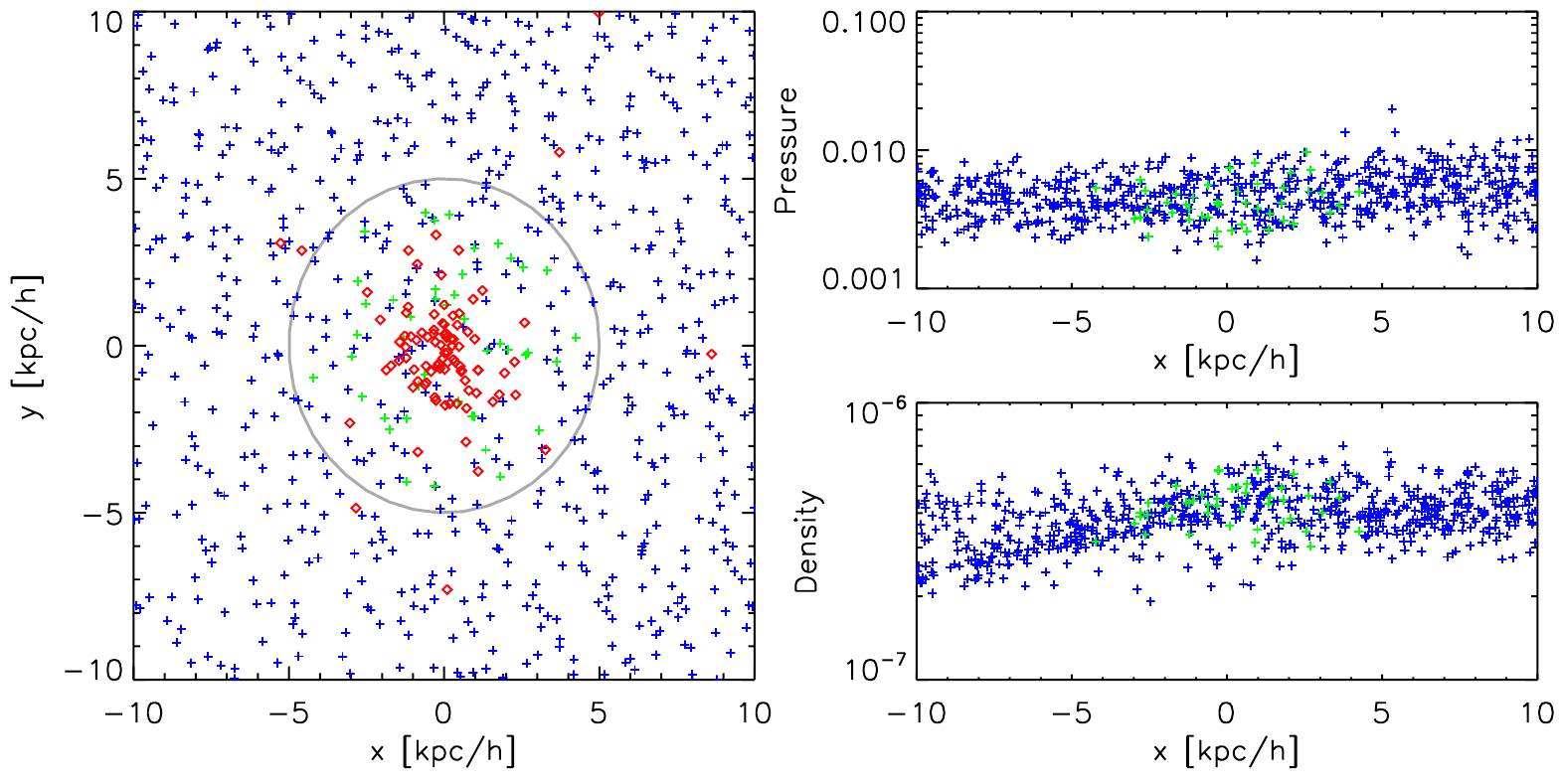}
  \caption{Gas and star particles for the same object as in
  \Fig{fig:ParticleComparison}. The left column shows all gas (blue)
  and star (red) particles inside a cubical region of side length
  10\hkpc\ as well as the gas particles inside a 5\hkpc\ sphere
  (green). The right column shows the pressure (upper) and density
  (lower) for the gas particles along the $x$-axis using the same
  colouring scheme.}
\label{fig:GasPressure}
\end{figure}

To verify that in actuality all gas particles should have been removed we
finally show in \Fig{fig:GasPressure} the same region again (left
panel, slightly different projection) alongside the pressure and
density of the gas particles along the $x$-axis (right panel). We can
clearly see that the gas particles inside the 5\hkpc\ sphere about the
centre (marked in green) are forming part of the overall background
gas particles (marked in blue). The stars (red) are only shown as a
reference.

In summary, while configuration space finders such as \ahf\ and
\subfind\ require a proper treatment of the gas' thermal energy to
remove (unbound) gas particles belonging to the background,
phase-/velocity-space based finders such as \rockstar\ and \stf\
encode part of this removal into their methodology. However, they keep
a residual amount of gas.

\section{Summary \& Conclusions}\label{sec:conclusion}
We set out to compare the performance of object finders for
cosmological simulations in the case where the data not only contains
dark matter but also baryons, i.e. gas and stars. Both types of matter
obviously cluster more strongly, gas being able to dissipate its
energy, and star particles forming at the bottom of deep potential
wells out of the cold gas. While star particles only feel
gravity (like the dark matter particles), they nevertheless originated
from different sites and more compact regions, respectively. Also
gas particles carry a thermal energy acting like a pressure (given
an appropriate equation of state). At present, halo finders deal with
these subtleties differently, with some properly accounting for the
gas pressure and others ignoring it. Therefore, the
fundamental question is, what will different halo finders find when it
comes to simulations including baryonic physics?

We found remarkable and excellent agreement between different halo
finders when applied to a constrained simulation of the Local Group
that includes all the relevant baryonic physics; at least as far as
dark matter and stellar content are concerned. However, there appear
to be strong discrepancies when it comes to the gas and hence baryon
fractions. This result may not be too surprising as both dark matter
and stars are particles primarily only feeling gravity (even though
stars obviously have a different formation history) whereas gas is
also subjected to hydrodynamical forces. It appears that -- to
properly identify and subsequently analyse and model the gas component
in cosmological simulations -- halo finders need to take into account
the pressure of the gas in the unbinding procedure, especially when
dealing with subhaloes containing of order 10,000 or less particles (in
total). But we have also seen that some correct accounting of the gas
happens in phase-/velocity-space based finders such as \rockstar\ and \stf\
used here: not all the gas particles inside the radius of the subhalo
are considered part of the object; only a residual number of gas
particles are kept. In these finders the (hot) gas component is prone
to be removed as they inherently use velocity information when
grouping and collecting the initial set of particles, whereas
configuration space finders only deal with velocities (either kinetic
or thermal) in a (post-processing) unbinding procedure.

In summary, the proper treatment of the gas certainly also depends on
the actual scientific question to be addressed with the numerical
data; any algorithm used for collecting gas particles should carefully
consider what type of study they are interested in. For instance,
for studies where the X-ray emission from (hot) gas in galaxy clusters is
sought, one clearly would not want to remove the majority of the gas
(during the unbinding or even before). Further, the differences will
certainly become more important for simulations (of subhaloes) with
much higher resolution where we expect subhaloes to hang on to their
own gas; we have seen here that for this particular data only a few
subhaloes still have gas (cf. \Fig{fig:MstarGasFunc}). But consider
gas rich objects at high redshift: for these the baryon fraction would
be dominated by the treatment of the gas and is expected to be much
larger as stars have not formed yet. This discussion about the
proper treatment of the gas (during the unbinding procedure) is
 akin to debates about unbinding in general: some
investigations heavily rely on it (e.g. subhalo dynamics) whereas
others can live without it (e.g. gravitational lensing). In any case,
the differences for the three main galaxies investigated here were
reassuringly negligible and we conclude that despite subtleties there is
in general a fair agreement of "galaxy finders".

\section*{Acknowledgements}
The work in this paper was initiated at the ``Subhaloes going Notts''
workshop in Dovedale, UK, which was funded by the European Commissions
Framework Programme 7, through the Marie Curie Initial Training
Network Cosmo-Comp (PITN-GA-2009-238356). We basically thank all the
participants of that meeting for all the stimulating discussions and
the great time in general.

AK is supported by the {\it Spanish Ministerio de Ciencia e
Innovaci\'on} (MICINN) in Spain through the Ramon y Cajal programme as
well as the grants AYA 2009-13875-C03-02, AYA2009-12792-C03-03,
CSD2009-00064, and CAM S2009/ESP-1496. He further thanks Christine
Corbett Moran for stimulating discussions about halo matching and
Marden Hill for the lost weekend. NIL acknowledges a grant from the DFG. PSB was supported by program number HST-AR-12159.01-A, provided by NASA through a grant from the Space
Telescope Science Institute, which is operated by the Association of
Universities for Research in Astronomy, Incorporated, under NASA
contract NAS5-26555. PJE acknowledges financial support from the
Chinese Academy of Sciences (CAS), from NSFC grants (No. 11121062,
10878001,11033006), and by the CAS/SAFEA International Partnership
Program for Creative Research Teams (KJCX2-YW-T23). KD acknowledges the support by the DFG Priority Programme 1177 and additional support by the DFG Cluster of Excellence "Origin and Structure of the Universe". HL acknowledges a fellowship from the European CommissionÕs Framework Programme 7, through the Marie Curie Initial Training Network CosmoComp (PITN-GA-2009-238356).

We like to thank the CLUES team (in particular Stefan Gottl\"ober,
Yehuda Hoffman, and Gustavo Yepes) for granting us access to the
simulation used for this comparison project. We also thank Andrea
Maccio for kindly providing us with the data of the observed
luminosity function (average of MW and M31). And last but not least we thank Volker Springel
for useful comments and remarks during the preparation of this work.

\bibliographystyle{mn2e}
\bibliography{archive}

\bsp

\label{lastpage}

\end{document}